\documentclass[prd,twocolumn,showpacs,nofootinbib,amsmath,amssymb,floatfix,superscriptaddress,showkeys]{revtex4}
\usepackage{multirow}
\usepackage{graphicx}% Include figure files
\usepackage{epstopdf}
\usepackage{dcolumn}% Align table columns on decimal point
\usepackage{bm}% bold math
\usepackage{threeparttable}
\usepackage{subfigure}
\usepackage{color,txfonts}
\usepackage{ulem}
\usepackage{makecell}
\definecolor{blue0}{rgb}{0,0,0.6}
\usepackage[colorlinks,linkcolor=blue0,anchorcolor=blue0,citecolor=blue0,urlcolor=blue0]{hyperref}

\newcommand{\beq}{\begin{equation}}
\newcommand{\eeq}{\end{equation}}
\newcommand{\beqa}{\begin{eqnarray}}
\newcommand{\eeqa}{\end{eqnarray}}

\begin{document}

\title{Investigating the correlations between IceCube high-energy neutrinos and Fermi-LAT $\gamma$-ray observations}

\author{Rong-Lan Li}
\affiliation{Laboratory for Relativistic Astrophysics, Department of Physics, Guangxi University, Nanning 530004, China}
\author{Ben-Yang Zhu}
\affiliation{Laboratory for Relativistic Astrophysics, Department of Physics, Guangxi University, Nanning 530004, China}
\author{Yun-Feng Liang}
\email[]{liangyf@gxu.edu.cn}
\affiliation{Laboratory for Relativistic Astrophysics, Department of Physics, Guangxi University, Nanning 530004, China}

\date{\today}

\begin{abstract}
We use 10 years of publicly available IceCube data to investigate the correlations between high-energy neutrinos and various Fermi-LAT gamma-ray samples. 
This work considers the following gamma-ray samples: the third Fermi-LAT catalog of high-energy sources (3FHL), $>100$ GeV Fermi-LAT events, LAT 12-year source catalog (4FGL), the fourth catalog of active galactic nuclei (4LAC) and subsets of these samples.
For each sample, both a single-source analysis and a stacking analysis are performed. We find no indication that the sources in these samples produce significant high-energy neutrinos.
From the null search results, we infer that each source population can produce no more than $\sim$2.5\%-36\% (at the 95\% confidence level, for a spectral index of $-2.5$) of the IceCube's diffuse neutrino flux.
Since we are using a larger (10 years) dataset of IceCube neutrinos, the constraints are improved by a factor of $\sim$2 compared to those based on 3 years of data.
\end{abstract}
%\pacs{95.35.+d, 95.85.Pw, 98.52.Wz}

\maketitle
\section{Introduction}
\label{sec_introduction}

The TeV-PeV diffuse astrophysical neutrino emission detected by IceCube~\cite{Aartsen:2013jdh, IceCube:2014stg, Aartsen:2015knd, IceCube:2016umi, Aartsen:2020aqd, Abbasi:2020jmh, IceCube:2021uhz} opens a new window for astrophysics~\cite{Kistler:2006hp, Beacom:2007yu, Murase:2010cu, Murase:2013ffa, Murase:2013rfa, Ahlers:2014ioa, Tamborra:2014xia, Murase:2014foa, Bechtol:2015uqb, Kistler:2016ask, Bartos:2016wud, Sudoh:2018ana, Bartos:2018jco, Bustamante:2019sdb, Hovatta:2020lor} and particle physcis~\cite{Beacom:2002vi,GonzalezGarcia:2005xw,Hooper:2005jp,Ioka:2014kca,Ng:2014pca,Aartsen:2017kpd,Bustamante:2017xuy,Zhou:2019vxt,Zhou:2019frk,Zhou:2021xuh,IceCube:2021rpz}. The origin of these neutrinos, however, remains a mystery. Great efforts have been made to find out the sources of these astrophysical neutrinos \cite{Abbasi:2010rd,Aartsen:2013uuv,Aartsen:2014cva,ANTARES:2015moa,Aartsen:2016oji,IceCube:2017der,Aartsen:2018ywr,Aartsen:2019fau,IceCube:2019lzm,IceCube:2020svz}. The two most likely TeV -- PeV neutrino emmiters are TXS 0506+056 and NGC 1068~\cite{IceCube:2018dnn,IceCube:2018cha,Aartsen:2019fau}. The TXS 0506+056 is found to have a 3$\sigma$ (global significance) correlation with a $\sim$300 TeV IceCube neutrino event~\cite{IceCube:2018dnn}. Furthermore, a neutrino flare was detected in the  direction of TXS 0506+056 between 2014 and 2015, with a global confidence of 3.5$\sigma$~\cite{IceCube:2018cha}. These two measurements are statistically independent, and a simple combination of the two results yields a significance of $\sim4.8\sigma$. However, multi-wavelength observations seem to not favor the possibility that the neutrinos are from TXS 0506+056 in a one-zone model, although more complicated two-zone models can solve this problem \cite{Keivani:2018rnh,Liu:2018utd,Gao:2018mnu,Xue:2019txw,Xue:2020kuw}. In a time-integrated analysis using 10 years of IceCube data, radio galaxy NGC 1068 is found to be a potential neutrino point source with a significance greater than 2.9$\sigma$ (global significance)~\cite{Aartsen:2019fau}.

Except the two above-mentioned sources (other less significant but promising candidates include: PKS B1424–418 \cite{Kadler:2016ygj}, PKS 1424+240 \cite{Padovani:2022wjk,Aartsen:2019fau}, PKS 1502+106 \cite{Rodrigues:2020fbu}, etc), searches for point sources in IceCube data always lead to only upper limits. These two sources contribute only a small fraction of the total flux of the diffuse neutrino emission. The origin of the majority of the diffuse astrophysical neutrinos is still unknown. The high-energy neutrinos observed by IceCube are likely generated by a large number of extragalactic sources. Many astrophysical sources have been proposed as possible sources of the high-energy neutrinos, including gamma-ray bursts~\cite{Waxman:1997ti,Abbasi:2009ig,2012ApJ...752...29H,Aartsen:2014aqy}, star-forming galaxies and starburst galaxies~\cite{Loeb:2006tw,He:2013cqa,Lunardini:2019zcf}, blazars and non-Blazar active galactic nuclei (AGNs)~\cite{Stecker:1991vm,Halzen:1997hw,Atoyan:2001ey}, tidal disruption events \cite{Wang:2015mmh}, etc. However, searches towards these sources do not show a strong correlation, they therefore are ruled out as the only primary sources of the IceCube diffuse neutrino flux. For example, the IceCube data from the directions of gamma-ray blazars have been analyzed and this type of sources are found to contribute at most 15\% to the diffuse neutrino flux~\cite{IceCube:2016qvd,Hooper:2018wyk,Smith:2020oac,Yuan:2019ucv}.
The possible association between radio bright active galactic nuclei and IceCube neutrino events has also been considered~\cite{Plavin:2020emb,Plavin:2020mkf}. 
Zhou et al.~\cite{Zhou:2021rhl} revisited the correlation between these radio-bright AGN and the TeV–PeV astrophysical neutrinos and found no strong correlation, indicating that no more than 30\% of the IceCube neutrino flux could be contributed by these objects.

In this paper, we focus on investigating the correlations between IceCube neutrinos and Fermi-LAT gamma-ray observations. {Some previous analyses on Fermi-LAT catalogs using IceCube data include: Refs.~\cite{IceCube:2016qvd,Hooper:2018wyk,Huber:2019lrm,luozhang2020,Smith:2020oac,IceCube:2021slf,IceCube:2022zbd}.} We search for evidence of neutrino emission in the directions of  Fermi-LAT sources of various catalogs and constrain their contributions to the IceCube diffuse neutrino flux. Our motivation is that, astrophysical neutrinos are expected to be produced by a hadronic process, so high-energy gamma rays will be emitted accompanied with the production of neutrinos. Since TeV photons are strongly absorbed by the extragalactic background light, for extragalactic sources the GeV gamma rays observed by Fermi-LAT are the messengers we can receive that are most close to IceCube's $>$100 TeV neutrinos in energy. As a result, the two are therefore more likely to correlate. 

This work utilizes a larger data set of IceCube neutrinos than previous similar analyses \cite{IceCube:2016qvd,Hooper:2018wyk,Smith:2020oac}. The latest (10 year, April 2008 to July 2018) public release of IceCube muon track data~\cite{IceCube:2021xar} are adopted. We will see that we can obtain stronger constraints. No potential neutrino point source is found, and no statistically significant correlation is found between IceCube events and any of considered Fermi-LAT samples. From the null searching results, we conclude that each of these source populations contribute no more than $\sim$0.8\%-89\% of IceCube diffuse neutrino flux.

\section{IceCube data and Fermi-LAT samples}
\label{sec_data}
\subsection{IceCube neutrino data}
\label{sec_data_nus}
The IceCube Neutrino Observatory detects neutrinos by detecting the Cherenkov light emitted by relativistic secondary charged particles from neutrino interactions~\cite{Achterberg:2006md}.
In this work, we utilize IceCube's updated public data release of muon tracks\footnote{DOI: \url{http://doi.org/DOI:10.21234/sxvs-mt83}}~\cite{IceCube:2021xar}. This dataset consists of muon tracks observed by IceCube from April 2008 to July 2018. The same data have been used in IceCube’s 10-year time-integrated neutrino point-source search~\cite{IceCube:2019cia}. The data contain both through-going and starting tracks. The former is primarily due to muon neutrino with interaction taking place outside the detector, while the latter are events that start within the instrument.
In total, there are 1134450 muon-track events contained in the data set.

{Most of the events in the data set are atmospheric muons and neutrinos produced in cosmic-ray air shower interactions which consist of the main background for searches of astrophysical neutrinos \cite{IceCube:2016zyt,IceCube:2021xar}.
For events from the southern hemisphere, the atmospheric muons can penetrate the ice and reach the IceCube, leading to a background with event rate orders of magnitude higher than the expected rate of astrophysical neutrinos. To reduce the background of atmospheric muons, a more stringent event selection is applied for the data from the southern sky.  For events from the northern hemisphere, atmospheric muons are filtered by the Earth. Atmospheric neutrinos can reach the detector from both hemispheres. Compared to astrophysical neutrinos, atmospheric neutrinos have a softer spectrum which dominates at energies $<$100 TeV.}

We use all IC40 to IC86-VII data, with the numbers in the names corresponding to the number of installed detector strings.
For each event in the dataset, the arrival time ($t$), {reconstructed muon energy ($E_{\mu}$)}, direction ($\alpha$, $\delta$) and directional uncertainty ($\sigma$) are recorded. The data set also provide binned detector response functions (effective areas, smearing matrices) as a function of declination and neutrino energy that will be used in the data analysis. We consider all neutrinos with $10^\circ<|b|<87^\circ$.

\subsection{Fermi-LAT gamma-ray samples}
\label{sec_data_srcs}
The Large Area Telescope (LAT) on board Fermi satellite is a wide field-of-view (FOV) imaging gamma-ray telescope, which observes gamma-ray phtons in the energy range from $\sim30\,{\rm MeV}$ to $>300\,{\rm GeV}$ \cite{Fermi-LAT:2009ihh}.
The Fermi-LAT began its observations in 2008 and surveys the entire sky each day. Its observation time and FOV overlap with the 10-year IceCube data considerably. With more than 13 years of observation, a variety of source catalogs of Fermi-LAT have been compiled and released. To investigate the correlations between GeV gamma-ray sources and TeV-PeV neutrinos, we consider the following Fermi-LAT samples.

We mainly consider the sources contained in the Third Catalog of Hard Fermi-LAT Sources (3FHL sample)\footnote{\url{https://fermi.gsfc.nasa.gov/ssc/data/access/lat/3FHL/}}~\cite{Fermi-LAT:2017sxy}. This catalog represents the hardest population of GeV gamma-ray sources, which are therefore more likely to be a hadronic origin. The 3FHL is constructed based on 7 years of Fermi-LAT data in the 10 GeV-2 TeV energy range~\cite{Fermi-LAT:2017sxy}. 
Besides the 3FHL, other samples considered in this work include, the gamma-ray sources of the fourth Fermi-LAT catalog (4FGL-DR3, for Data Release 3)\footnote{\url{https://fermi.gsfc.nasa.gov/ssc/data/access/lat/12yr_catalog/}}~\cite{Fermi-LAT:2019yla,Fermi-LAT:2022byn} and the fourth catalog of active galatic nuclei (4LAC-DR2, for Data Release 2)\footnote{\url{https://fermi.gsfc.nasa.gov/ssc/data/access/lat/4LACDR2/}}~\cite{Fermi-LAT:2019pir,Lott:2020wno}. All these samples are summarized in Table~\ref{tab:label}. For each sample, we exclude the sources categorized as pulsars (PSRs) from the catalog. To avoid the complexity of the Galactic Plane, we only select $\left | b \right |>10^{\circ}$ sources from the catalogs.

Additionally, we search for excess neutrino emissions in the directions of $>$100 GeV high-energy Fermi-LAT events (HEE sample). This sample is used as a proxy of emissions from all hard sources including those too faint to be resolved from the background. {The HEE events are extracted from the all-sky weekly Fermi-LAT data\footnote{\url{http://heasarc.gsfc.nasa.gov/FTP/fermi/data/lat/weekly/photon/}}.
We consider 12 years of Fermi-LAT data from 2008-08-04 to 2020-10-26  (MET 239557417-625393779). 
We select events of Pass 8 SOURCE event class ({\tt evclass=128, evtype=3}) with energies $>$1 GeV. 
Events with zenith angles $z_{\rm max}>90^\circ$ are removed to avoid the contamination from the Earth's limb. 
We also use the filter {\tt (DATA\_QUAL>0 \&\& LAT\_CONFIG==1)} to select events within good time intervals. After these selections, we obtain 13335 events of the HEE sample.}

\begin{table*}[!htbp]
	\centering
	\caption{Fermi-LAT samples considered in this work.}
	\begin{tabular}{p{3cm}p{2.5cm}p{2.5cm}p{2cm}p{3cm}p{1.5cm}}
		\hline
		\hline
		Catalog Name & Energy Range$^a$ & Time interval$^a$ & $N_{\rm src}^b$ & $f_{\rm UL}$ for $\Gamma=-2.5^c$ & TS$_{\rm max}$ \\
		\hline
		4FGL-DR3 & \multirow{4}{*}{50 MeV-1 TeV} & \multirow{4}{*}{12 years} & 4689 & 18.0\% \,\,\,/ 5.8\% & 0.00\\
		4FGL-DR3 Blazars &  &  & 3339 & 20.6\% \,\,\,/ 5.5\% &0.00 \\
		4FGL-DR3 BL Lacs &  &  & 1354 & 36.1\% / 9.9\% &2.38 \\
		4FGL-DR3 FSRQs &  &  & 757 & 6.7\% \,\,\,/ 2.5\% &0.02 \\
		\hline
		4LAC-DR2 & \multirow{4}{*}{50 MeV-1 TeV} & \multirow{4}{*}{10 years} & 3128 & 23.8\% \,\,\,/ 5.7\% &0.00 \\
		4LAC-DR2 Blazars &  &  & 3060 & 21.4\% \,\,\,/5.3\% &0.00 \\
		4LAC-DR2 BL Lacs &  &  & 1236 & 34.5\% / 9.9\% &2.29\\
		4LAC-DR2 FSRQs &  &  & 707 & 7.7\% \,\,\,/2.4\% &0.02\\
		\hline
		3FHL & \multirow{4}{*}{10 GeV-2 TeV} & \multirow{4}{*}{7 years} & 1215 & 21.6\%\,\,\, / 8.0\% &0.35\\
		3FHL Blazar &  &  & 1078 & 21.0\% \,\,\,/ 7.5\% &0.42\\
		3FHL BL Lacs &  &  & 697 & 25.4\% \,\,\,/ 6.6\% &2.13\\
		3FHL FSRQs &  &  & 165 & 5.2\% \,\,\,/ 3.6\% &0.01\\
		\hline
		HEE & {$>$100~GeV} & {12 years} & 13335 & 15.3\% \,\,\,/ 15.0\% &0.00\\
		\hline
		\hline
	\end{tabular}
\begin{tablenotes}
\item $^a$ The energy range and time interval of the Fermi-LAT data that used to construct the catalogs. \\
\item $^b$ Number of sources / events included in the sample. The sources categorized as pulsars or in the $|b|<10^{\circ}$ region have been excluded.\\
\item $^c$ Upper limit on the fraction of the IceCube diffuse neutrino flux that originates from the sources in the corresponding catalogs. The first (second) value is for equal weighting (flux weighting) scheme. {For the HEE sample, the second value uses the $N_{\rm events}$ in each pixel as the weighting.}
\end{tablenotes}
	\label{tab:label}
\end{table*}

\section{Analysis Method}
\label{sec_ana}
We perform an unbinned likelihood analysis on the neutrino data to derive the significance of neutrino signal from the direction of target source, using the neutrino events data together with the public instrument response functions of IceCube. We refer to Ref.~\cite{Braun:2008bg,Zhou:2021rhl} for the analysis method used in this work.

\subsection{Single-source analysis}
\label{sec_ana_single}
The unbinned likelihood for neutrino data is evaluated over all events within a region of interest (ROI), for which we choose a 5$^\circ$ circle surrounding the target. The likelihood reads
\begin{equation}
\mathcal{L}(\Phi_{0},\Gamma)=\prod_i\left[\frac{n_s(\Phi_{0},\Gamma)}{N}{S}_i+\left(1-\frac{n_s(\Phi_{0},\Gamma)}{N}\right){B}_i\right],
\label{eq:1}
\end{equation}
where $n_s$ and $N$ are the numbers of signal neutrinos and total observed neutrinos within the ROI, respectively. 
The ${S}$ (${B}$) is the signal (background) probability density function (PDF). For a generic likelihood analysis, the PDF is the product of independent space, energy and time probability terms. In this work we ignore the energy term of the PDF.
As is suggested in~\cite{Braun:2008bg}, the use of energy information would improve discovery potential by almost a factor of two. We leave the further analyses that take into account the energy PDF term for potential future work. Since we are doing a time-integrated analysis, we also neglect the time related PDF term. As a result, the PDFs contain only the spatial distribution. 

For the signal PDF, it is described by a two-dimensional Gaussian:
\begin{equation}
{S}_i(\theta_i,\sigma_i) = \frac{1}{2\pi\sigma_i^2}\exp(-\frac{\theta_i^2}{2\sigma_i^2}),
\label{eq:single_si}
\end{equation}
where $\theta_i$ is the angular separation to the source for the $i$-th neutrino in the ROI, and $\sigma_i$ is the directional reconstruction uncertainty of the neutrino event.
For background PDF, the slight declination ($\delta$) dependence of the atmospheric neutrino background can be ignored in a small ROI.
It is thus reasonable to assume the background neutrinos are uniformly distributed across the ROI, ${B_i}({\rm sin}\delta_i)={1}/{\Omega}$, where $\Omega$ is the solid angle of the $5^\circ$ ROI.

The $n_s$ in Eq.~(\ref{eq:1}) is the model-predicted number of signal neutrinos, given by
\begin{equation}
n_s=T\times\int A_{\rm eff}(E,{\rm sin}\delta_i)\Phi(E)dE
\label{eq:single_ns}
\end{equation}
for a total detector uptime of $T$ and a signal flux model of the form
\begin{equation}
\Phi(E)=\Phi_{0}(\frac{E}{100~{\rm TeV}})^{\Gamma}.
\label{eq:single_phi}
\end{equation}
The $A_{\rm eff}$ above is the  effective area of the detector in the declination of $\delta_i$.

By maximizing the likelihoods of Eq.~(\ref{eq:1}) for both signal model, $\mathcal{L}(n_s)$ (with $n_s$ free to vary), and background-only model, $\mathcal{L}(n_s=0)$, the test statistics (TS) of the signal can be derived by comparing the likelihood values between the two,
\begin{equation}
{\rm TS}=2{\rm log}\frac{\mathcal{L}(\hat{n}_s)}{\mathcal{L}(n_s=0)},
\label{eq:ts}
\end{equation}
where the $n_s$ with a hat ($\hat{n}_s$) denote the best-fit quantity. If the background hypothesis is true, the probability distribution for $\sqrt{\rm TS}$ is approximately a standard normal distribution (i.e., $\chi^2$ distribution with one degree of freedom)~\cite{Wilks:1938dza}. A significant deviation from the normal distrubution indicate the null hypothesis is rejected (i.e., the presence of excess neutrinos over the background).

\subsection{Stacking analysis}
\label{sec_ana_joint}
To improve sensitivity, all the sources in each Fermi-LAT sample are combined into a single stacking analysis. 
In such a case, the signal PDF is composed of the contributions from all sources (indexed by $j$):
\begin{equation}
S_i = \frac{\sum_j \omega_{j,{\rm model}}\omega_{j,{\rm acc}}S_{ij}}{\sum_j\omega_{j,{\rm model}}\omega_{j,{\rm acc}}},
\label{eq:si}
\end{equation}
where $S_{ij}$ is the signal PDF of single source and is just the above two-dimensional Gaussian PDF:
\begin{equation}
S_{ij} =  S(\vec{x_i},\sigma _i,\vec{x_j}) = \frac{1}{2\pi\sigma_i^2}\exp\left(-\frac{\theta(\vec{x_i}, \vec{x_j})^2}{2\sigma_i^2}\right)
\label{eq:sij}
\end{equation}
Here $\vec{x}_i$ and $\vec{x}_j$ are the directions of the neutrino event $i$ and source $j$, and $\theta(\vec{x}_i, \vec{y}_j)$ is the angular separation between the source and event.
The $\omega_j$ in Eq. (\ref{eq:si}) are weighting factors of source $j$. The spectral index $\Gamma$ of the signal neutrinos is assumed to be the same for all the sources in our analysis. Thus, the $\omega_{j,{\rm acc}}$ term is mainly related to the detector's acceptance,
\begin{equation}
\omega_{j,{\rm acc}}(\delta_j)=T\times\int A_{\rm eff}(E,\delta_j)E^{\Gamma}dE.
\end{equation}

For the $\omega_{j,{\rm model}}$ we consider two weighting schemes. 
The first one assumes that the high-energy neutrino flux of the source is independent of its gamma-ray flux, $\omega_{j,{\rm model}}=1$.
For the second scheme, we assume that the high-energy neutrino flux is proportional to the gamma-ray band flux $f_\gamma$ reported in the catalogs, $\omega_{j,{\rm model}}=f_\gamma$.
In this scheme, those sources brighter in the gamma-ray band contribute more to the signal PDF.

{The majority of the background  for astrophysical neutrinos  are muons and neutrinos produced from CRs interaction in the Earth's atmosphere.}
The background PDF is derived directly from IceCube neutrino data,
\begin{equation}
B_i(\delta_i)=\frac{N_{\delta_i\pm3}}{N_{\rm tot}\Omega_{\delta_i\pm3}},
\end{equation}
where $N_{\delta_i\pm3}$ and $N_{\rm tot}$ are the number of neutrino events in the $\delta_i\pm3^{\circ}$ region and the total number of all-sky events, respectively; $\Omega_{\delta_i\pm3}$ is the solid angle of the $\delta_i\pm3^{\circ}$ region.  

The 10-year IceCube data contains neutrino events from 10 data seasons: IC40, IC59, IC79, IC86-I, IC86-II to IC86-VII. The total likelihood combining 10 data samples is
\begin{equation}
\mathcal{L}(\Phi_{0},\Gamma)=\prod_k \mathcal{L}_k(\Phi_{0},\Gamma;T_k,A_{{\rm eff},k},\{\boldsymbol{\alpha},\boldsymbol{\delta},\boldsymbol{\sigma}\}_k),
\end{equation}
where $\mathcal{L}_k$ is the likelihood for each data sample $k$ calculated through the above-described procedure and $\{\boldsymbol{\alpha},\boldsymbol{\delta},\boldsymbol{\sigma}\}_k$ is the neutrino list of the data sample $k$.

\begin{figure*}[t]
\centering
\includegraphics[width=0.45\textwidth]{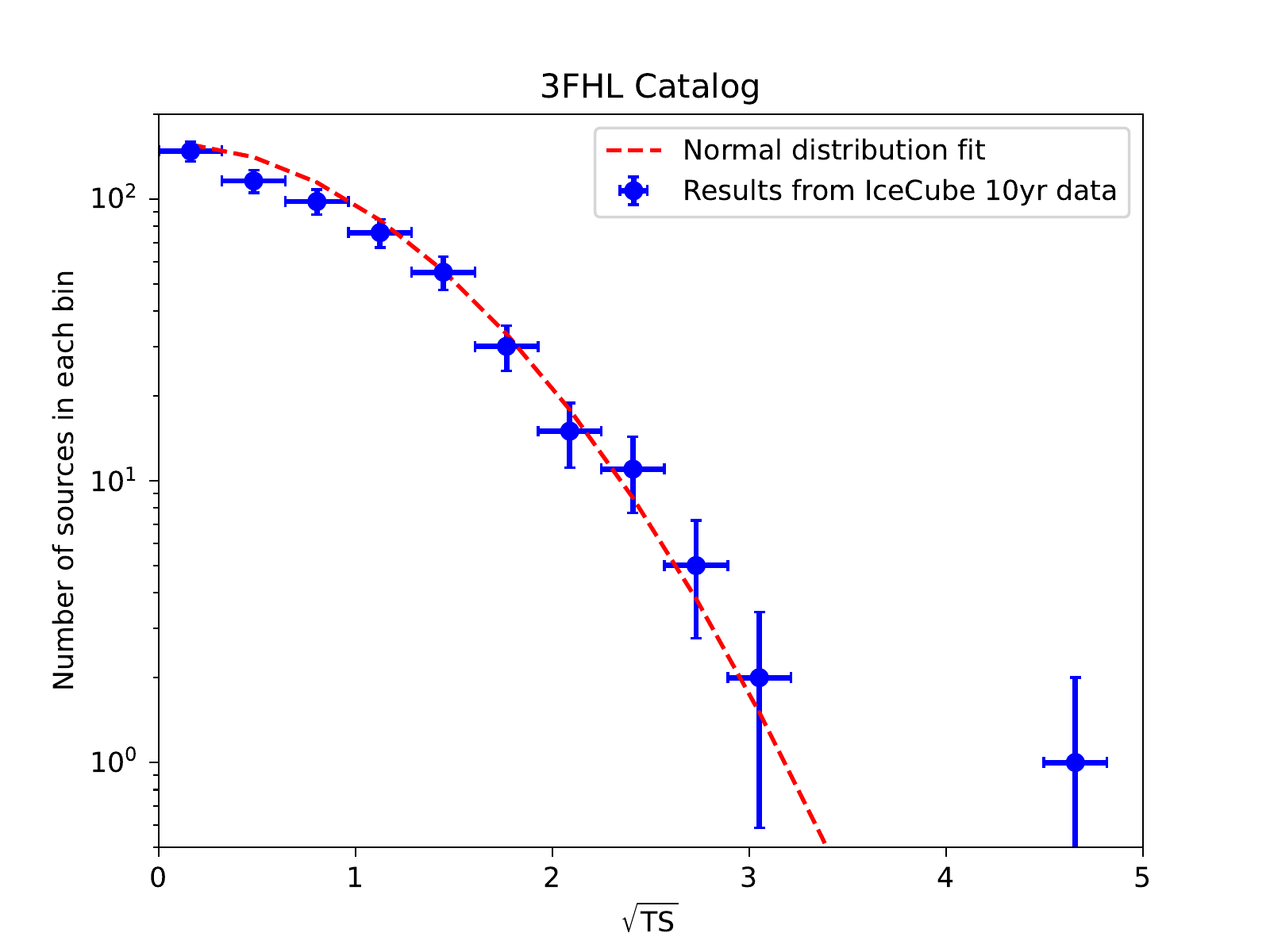}
\includegraphics[width=0.45\textwidth]{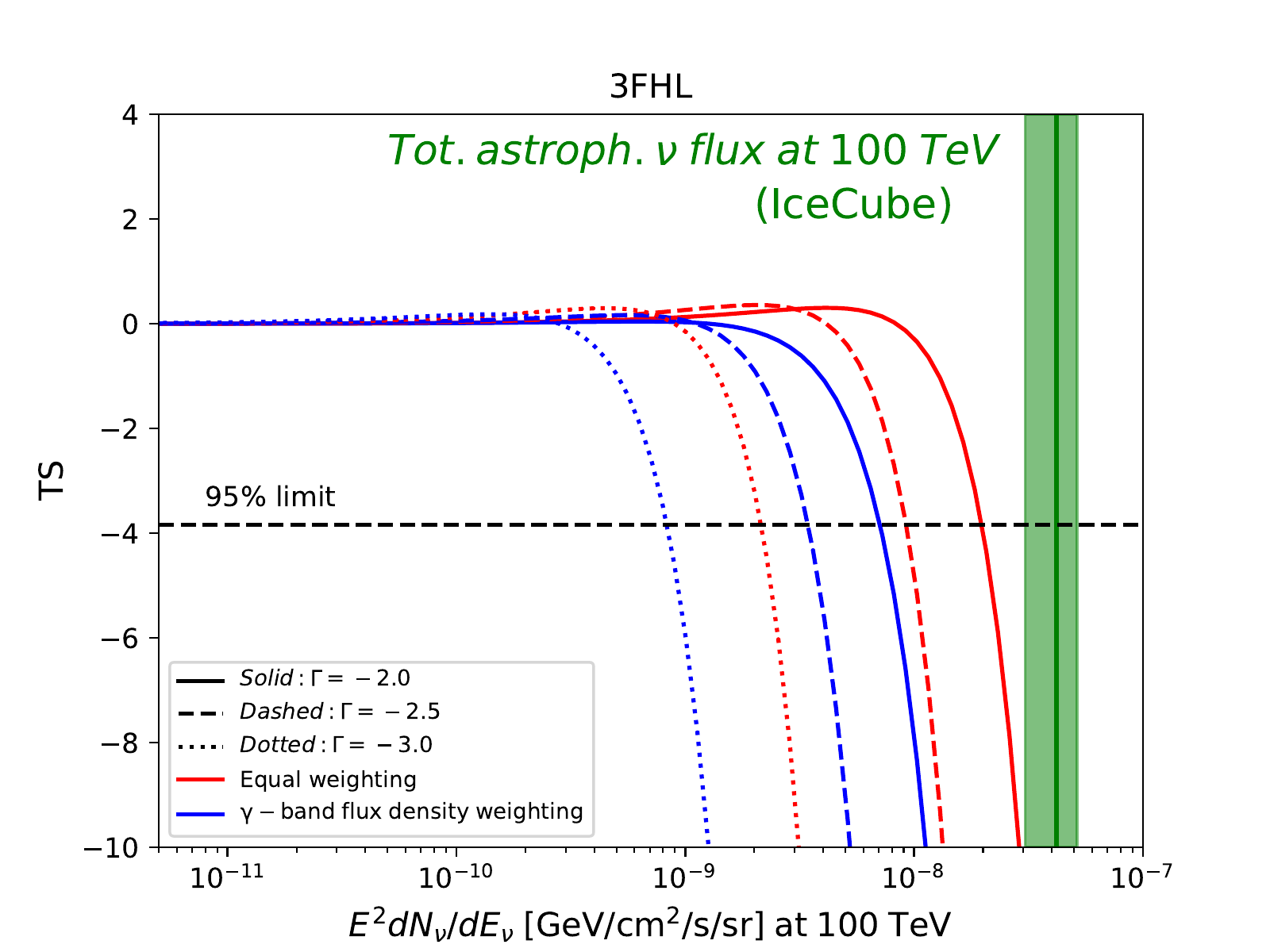}
\includegraphics[width=0.45\textwidth]{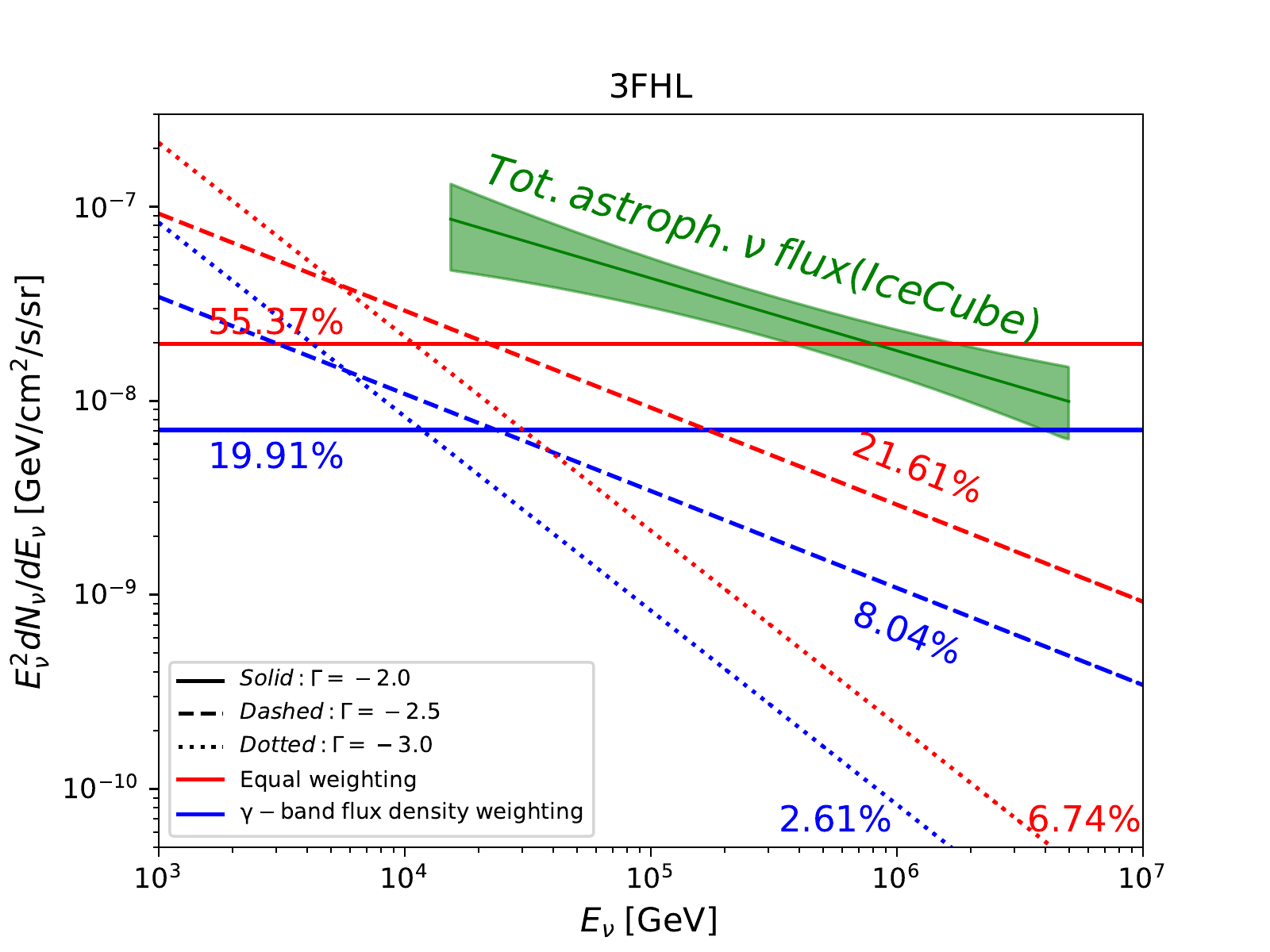}
\includegraphics[width=0.45\textwidth]{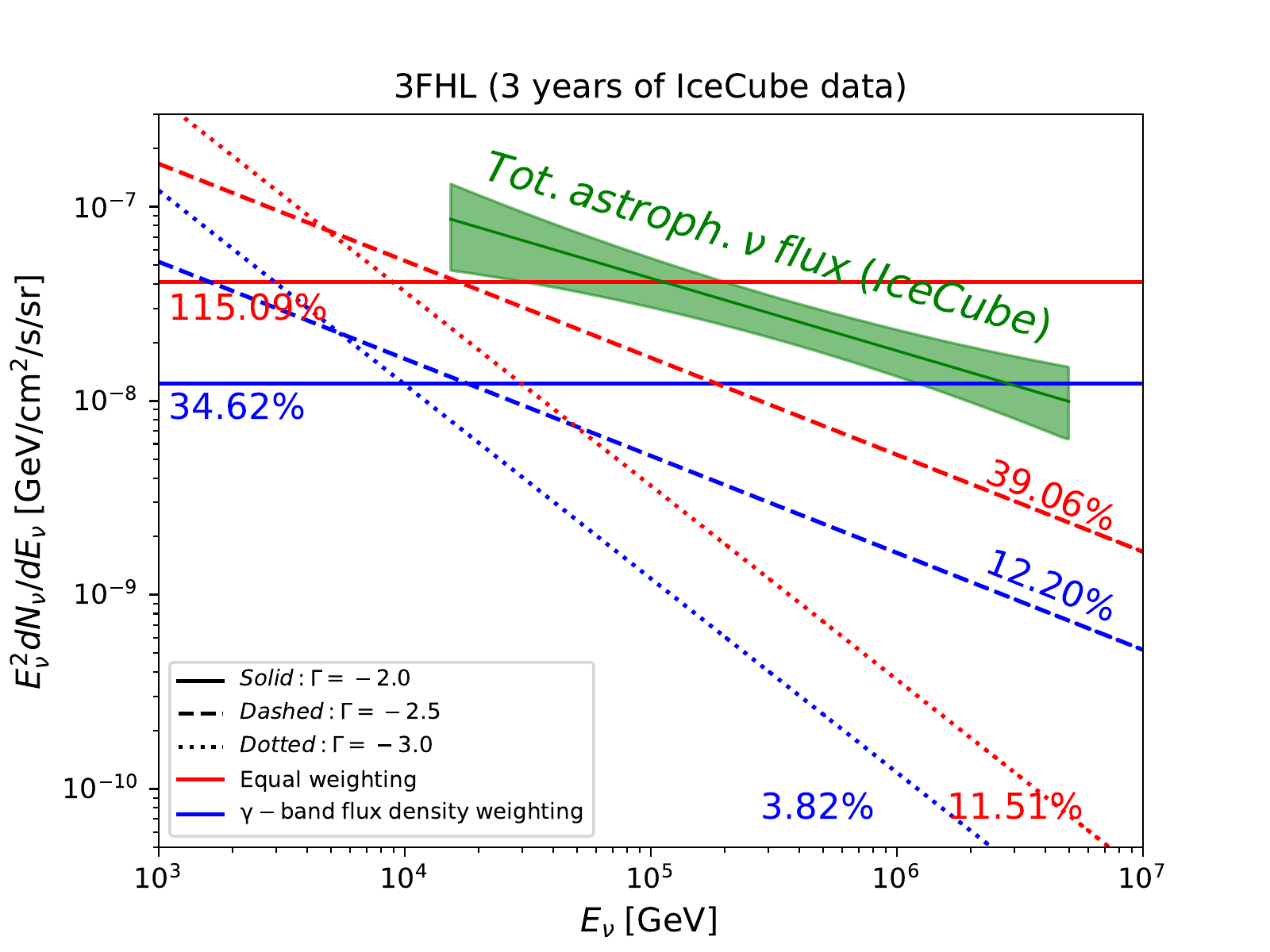}
\caption{{\it Upper left}: Distribution of TS values for the sources in the 3FHL with $\hat{n}_s>0$ from our single-source analysis with 10-year IceCube muon-track data. The error bars of the data points are given by the square root of the source count in each bin. The red line is the standard normal distribution which is the expectation of the background hypothesis. {\it Upper right}: The change of the TS value as a function of the total neutrino flux from the sources in the 3FHL catalog, for different choices of the neutrino spectral index (solid, dashed and dotted lines) and for two weighting schemes (red and blue lines). Also shown is the all-sky diffuse astrophysical neutrino flux measured by IceCube (vertical green band) \cite{IceCube:2021uhz}. {\it Lower left}: The 95\% confidence level upper limits on the total neutrino flux from the sources in the 3FHL sample. The constraints are compared to the all-sky diffuse neutrino flux measured by IceCube. The number above each line displays the maximum fraction that 3FHL sources can contribute to the diffuse neutrino flux. {\it Lower right}: The same as the lower-left panel but for results based on only IC79+IC86I+IC86II data.}
\label{fig_3fhl}
\end{figure*}

\section{Results}
\label{sec_results}
Using the method described in Section~\ref{sec_ana}, we search for the correlations between the neutrino data and the GeV observations from Fermi-LAT. 

\subsection{3FHL}

We first consider the sources included in the third Fermi-LAT catalog of high-energy sources (3FHL). The 3FHL represents a population of sources with relatively hard energy spectra, and according to the extrapolation of the spectrum, they are more likely to have observable neutrino emission. The catalog contains a total of 1497 gamma-ray sources, of which 1212 are classified as Blazars, including 172 flat-spectrum radio quasars (FSRQs) and 750 BL Lacs. In our analysis, we do not consider the sources in the region of $|b|<10^\circ$ to avoid the uncertainty from the Galactic Plane. The sources in the regions of $|{\rm Dec}|>87^\circ$ are excluded since the background PDF can not be reliably determined. We also remove PSRs from the sample because of their leptonic origin. The number of sources finally used for analysis is listed in Table~\ref{tab:label}.

For the 1215 sources in the 3FHL sample, we first perform a single-source analysis to search for individual neutrino point sources in the IceCube data. In the upper left panel of Fig.~\ref{fig_3fhl}, for 1215 sky locations associated with 3FHL sources, we show the distribution of $\sqrt{\rm TS}$ value given by the analysis. 
We compare this histogram with a standard normal distribution (red dashed line), which is an expectation from Gaussian fluctuations of the background. It can be seen that for the 3FHL sources the $\sqrt{\rm TS}$ distribution generally follows the standard normal distribution. 
We note that there is one source showing relatively high significance of TS$\sim$22, which seems deviating from the distribution of background fluctuations. After examination, we find that the source is NGC 1068. The presence of a possible neutrino excess in the direction of NGC 1068 has already been reported by the IceCube collaboration \cite{Aartsen:2019fau}. The result of NGC 1068 therefore suggest that our analysis is reliable. 
Except for NGC 1068, we observe no significant evidence of a deviation from Gaussian fluctuations. According to the Wilks’ theorem~\cite{Wilks:1938dza}, this means that the background hypothesis is favored and there is no strong correlation between the diffuse astrophysical neutrinos and the 3FHL catalog.

Next, we perform a stacking analysis of all sources of 3FHL sample to investigate the contribution of such source population to the diffuse neutrino flux observed by IceCube. In the upper right panel of Fig.~\ref{fig_3fhl}, we show the TS profile as a function of the total neutrino flux from 3FHL sources for the two weighting schemes described in Section~\ref{sec_ana_joint}. 
{The $\nu_\mu+\bar{\nu}_\mu$ flux relevant to the analysis of muon track data is converted to all six-flavor neutrino flux by multiplying a factor of 3 (i.e., assuming equal flux for all six flavors).} 
Considering that the diffuse astrophysical neutrinos measured by the IceCube has a spectral index of $\sim-2.3$ to $-2.7$ \cite{IceCube:2021uhz}, we take three different spectral indices ($\Gamma=-2.0, -2.5$ and $-3.0$) in the anslysis. As shown, none of the cases reveals a significant correlation between the sample and IceCube neutrinos. Thus, we place upper limits on the total neutrino emission contributed by these 3FHL sources at a 95\% confidence level (corresponding to the horizon line of ${\rm TS}\simeq-3.84$). The limits we obtain are lower than the diffuse flux reported by the IceCube collaboration, indicating that these sources cannot produce all the observed astrophysical neutrinos.

In the lower left panel of Fig.~\ref{fig_3fhl}, we present the upper limits of neutrino flux derived in our analysis and compare it with IceCube’s diffuse neutrino flux. {Here, we adopt the the most up-to-date muon–neutrino flux reported in Ref.~\cite{IceCube:2021uhz} for comparison, which is based on 9.5 years of track data and best fitted with a power law of $dF/dE_\nu=\Phi_0\times (E_\nu/100\,\rm{TeV})^{-2.37}$, where $\Phi_0$ is the flux normalization.} We find that the 3FHL sample can contribute no more than 55.37\% of IceCube's diffuse neutrino flux ($\int E_\nu dF/dE_\nu\, dE_\nu$). 
In the energy range of 16 TeV-2.6 PeV, this catalog accounts for at most 55.37\% ($\Gamma=-2.0$), 21.61\% ($\Gamma=-2.5$) and 6.74\% ($\Gamma=-3.0$) for equal weighting and at most 19.91\% ($\Gamma=-2.0$), 8.04\% ($\Gamma=-2.5$) and 2.61\% ($\Gamma=-3.0$) for flux weighting.

This work uses the latest 10 years of IceCube neutrino data, so the constraints are effectively improved due to the longer data set used. We have tested that if repeating our analysis with only 3 years of data (IC79, IC86-I, IC86-II), the obtained upper limits would be weakened by a factor of $\sim$2 (see lower right panel of Fig.~\ref{fig_3fhl}). Since there is currently no similar work (especially focusing on 3FHL sources) based on the 10-year data, the constraints presented here on the contribution from 3FHL sources would be by far the strongest one.

\begin{figure*}[t]
\centering
\includegraphics[width=0.45\textwidth]{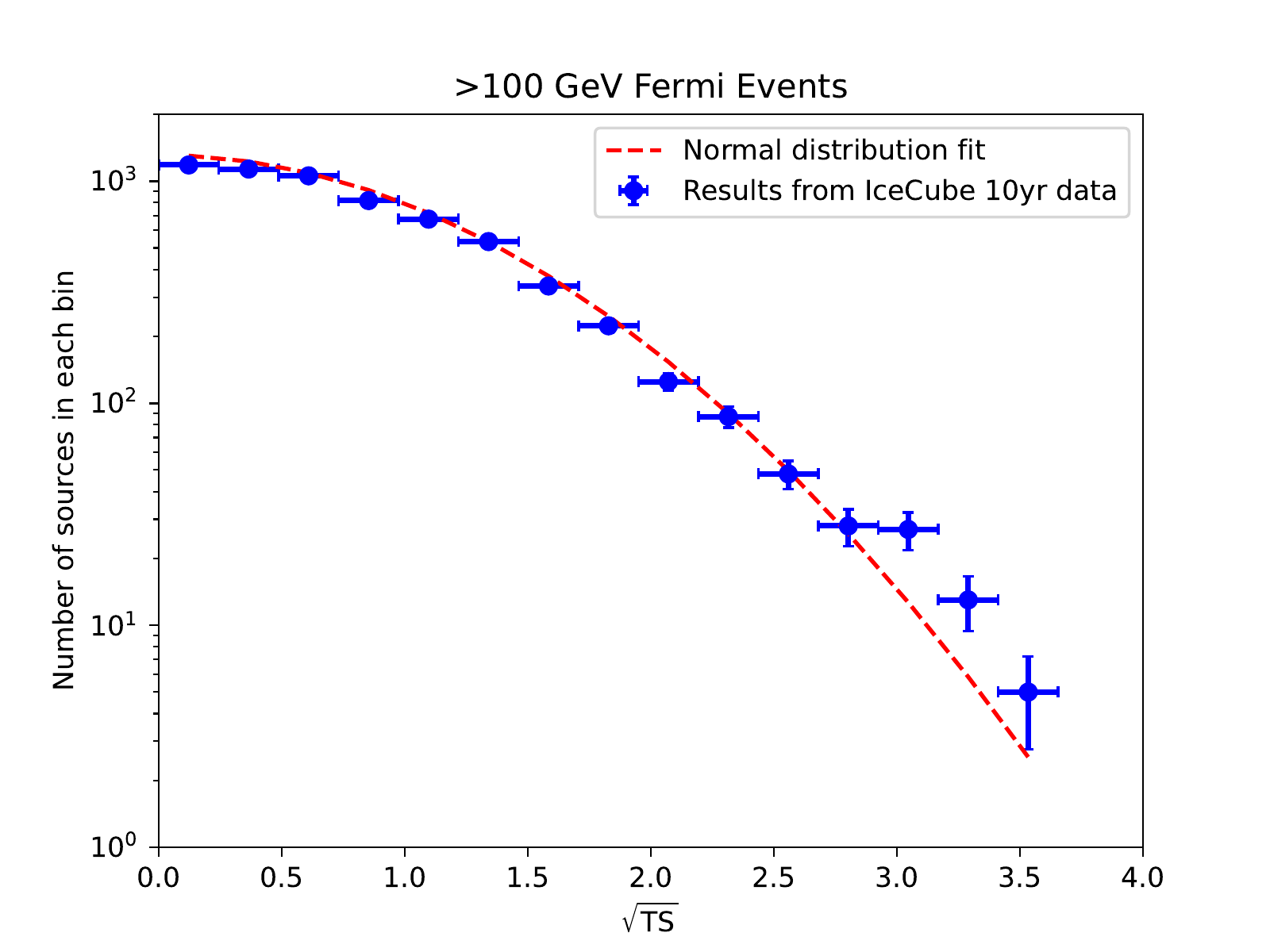}
\includegraphics[width=0.45\textwidth]{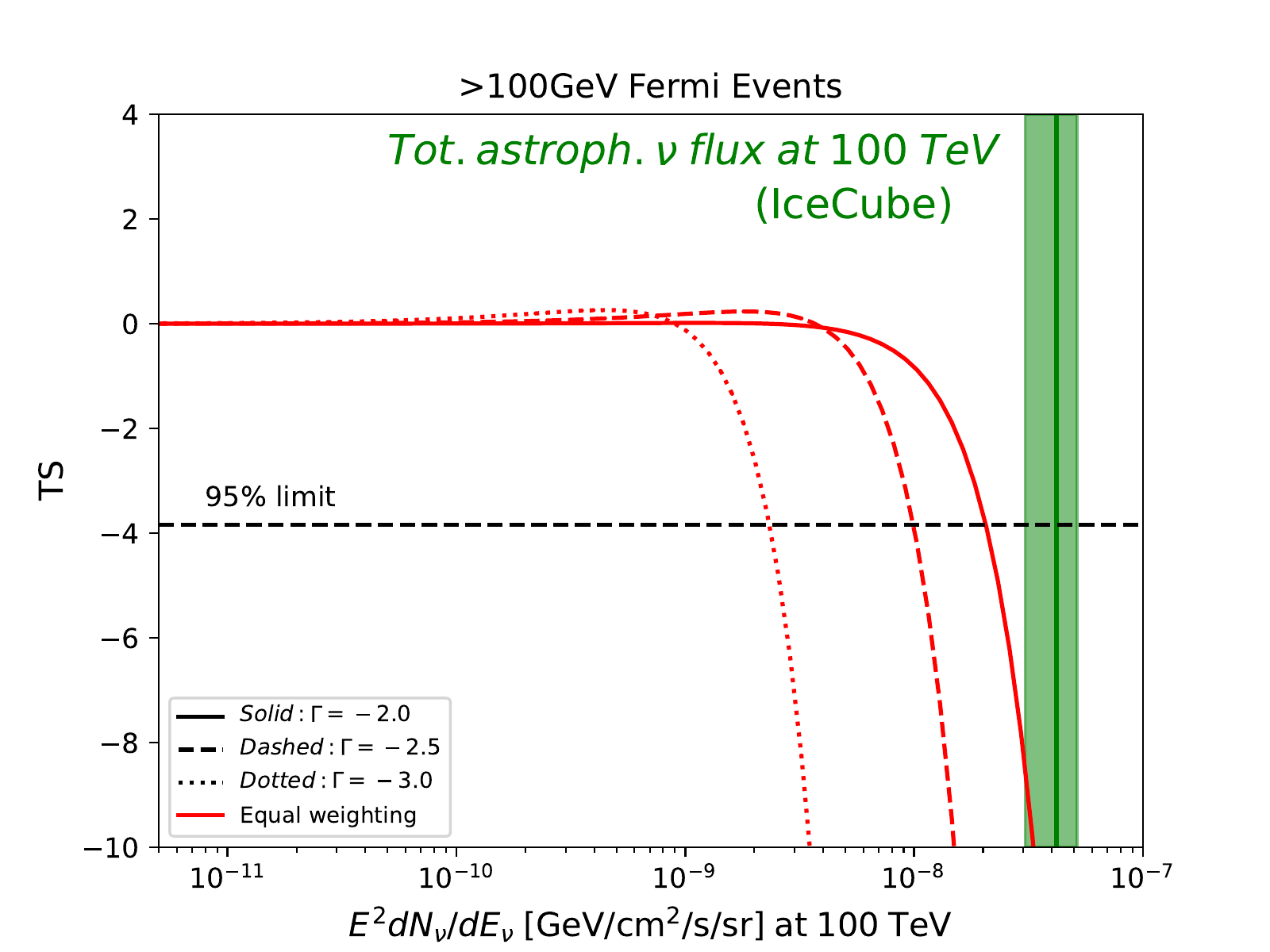}
\caption{{\it Left}: Distribution of TS values for putative point sources in the directions of Fermi-LAT events with $E_\gamma>100$ GeV and with $|b|>10^\circ$. The red line is the standard normal distribution which is the expectation of a background hypothesis. {\it Right}: Change of the TS value as a function of the total neutrino flux from the directions of Fermi-LAT high-energy events.
}
\label{fig_100}
\end{figure*}

\subsection{3FHL blazars}
\label{sec_res_3fhl}
The tentative neutrino excess observed in the direction of TXS 0506+056 provide evidence in favor of blazars as sources of high-energy neutrinos. We next separately consider the gamma-ray sources identified as blazars in the 3FHL catalog. We also further divide the blazars into two subclasses to perform the analysis: flat spectrum radio quasars (FSRQs) and BL Lacertae objects (BL Lacs). 
Again, we find no evidence to support neutrino emission from these source populations and place strong constraints on their contributions to IceCube's diffuse neutrino flux. We find that the total 3FHL blazars/BL LACs/FSRQs can contribute at most 54.53\%/60.28\%/15.51\% of the diffuse flux (see Table~\ref{tab:label} for the results of $\Gamma=-2.5$, and related figures are shown in the Appendix). 
{As a comparison, the IceCube Collaboration has also performed a stacking analysis to search for neutrino emission from 3FHL blazars with 8 years of IceCube northern-sky data \cite{Huber:2019lrm}. They reported an 90\% upper limit of 13.0\%-16.7\% on the contribution from 3FHL blazar population for $\Gamma=-2$, consistent with the 18.8\% upper limit (95\% C.L.) obtained in this work for the flux weighting scheme. }

Note that our results here consider only the blazars within the 3FHL catalog. The contributions from those blazars that are too far away or too weak to be included in the catalog are not taken into account. To translate the results to apply to all blazars in the Universe, we need to multiply a completeness factor of ~1-2 \cite{Hooper:2018wyk,Smith:2020oac}, which would not change the results substantially.

We notice that of all four 3FHL samples (all, blazar, BL Lac, FSRQ), the analysis of BL Lacs gives almost the weakest upper limits on the contribution to the diffuse neutrino flux (especially for the equal weighting scheme), although the BL Lac sample does not contain the largest number of sources (see Table~\ref{tab:label}). Similar results are also revealed (and are more evident) in the below analysis of 4FGL sources (see Section~\ref{sec_res_4fgl4lac}). In addition, in our stacking analysis, the search for neutrino excess from BL Lacs gives the only non-negative TS value (TS$\sim$2.3). Note that the 3FHL source catalog does not contain the BL Lac source TXS 0506+056 (the exclusion of TXS 0506+056 in the analysis of the 4FGL sources has a negligible effect on the results). Similar indication has also been discussed in Ref.~\cite{Smith:2020oac}. All these results support that the BL Lac blazars may have higher contribution to the IceCube's diffuse neutrinos. However, a statistically significant evidence is still lacking in current analysis.

\subsection{Very-high energy Fermi events (HEE)}
\label{sec_res_hee}
According to our analysis above (as well as previous works \cite{IceCube:2016qvd,Hooper:2018wyk,Smith:2020oac}), the diffuse emission of high-energy neutrinos observed by IceCube cannot be entirely accounted for by a small number of bright sources. Only a fraction (no more than 1.2-60.3\%, depending on the source populations considered and the different analysis configurations) of the neutrino flux may originate from the resolved 3FHL sources. Most of the astrophysical neutrinos must have been produced by a large number of faint sources. These sources are obscured and cannot be resolved from the gamma-ray background. But it is a reasonable speculation that, in general there will be a higher probability to detect high-energy gamma rays in the directions of neutrino sources. 
On the other hand, the 3FHL catalog is constructed based on gamma-ray observations of $>10\,{\rm GeV}$, while it is possible that the gamma-ray emission from TeV-PeV neutrino sources dominates at even higher energies (e.g., $>100$~GeV). So we directly examine the correlation between IceCube muon-track events and high-energy Fermi-LAT photons. 
More specifically, we look for neutrino emission in the directions of $>100\,{\rm GeV}$ Fermi-LAT events with the method of Section~\ref{sec_ana}. To avoid contamination from the Galactic Plane, we only use the events with $|b|>10^\circ$, which includes 13335 photon events. 

\begin{figure*}[tb]
\centering
\includegraphics[width=0.45\textwidth]{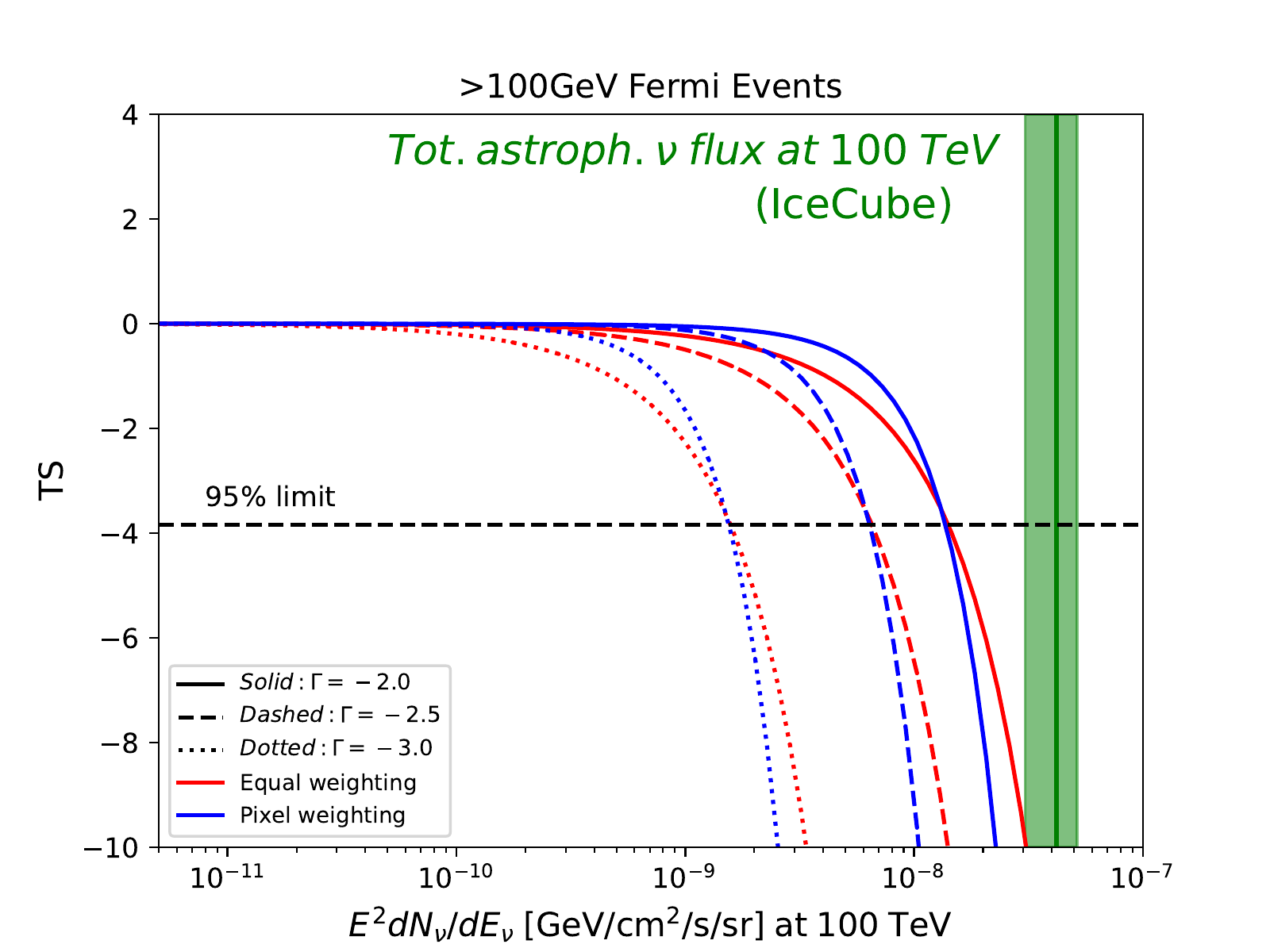}
\includegraphics[width=0.45\textwidth]{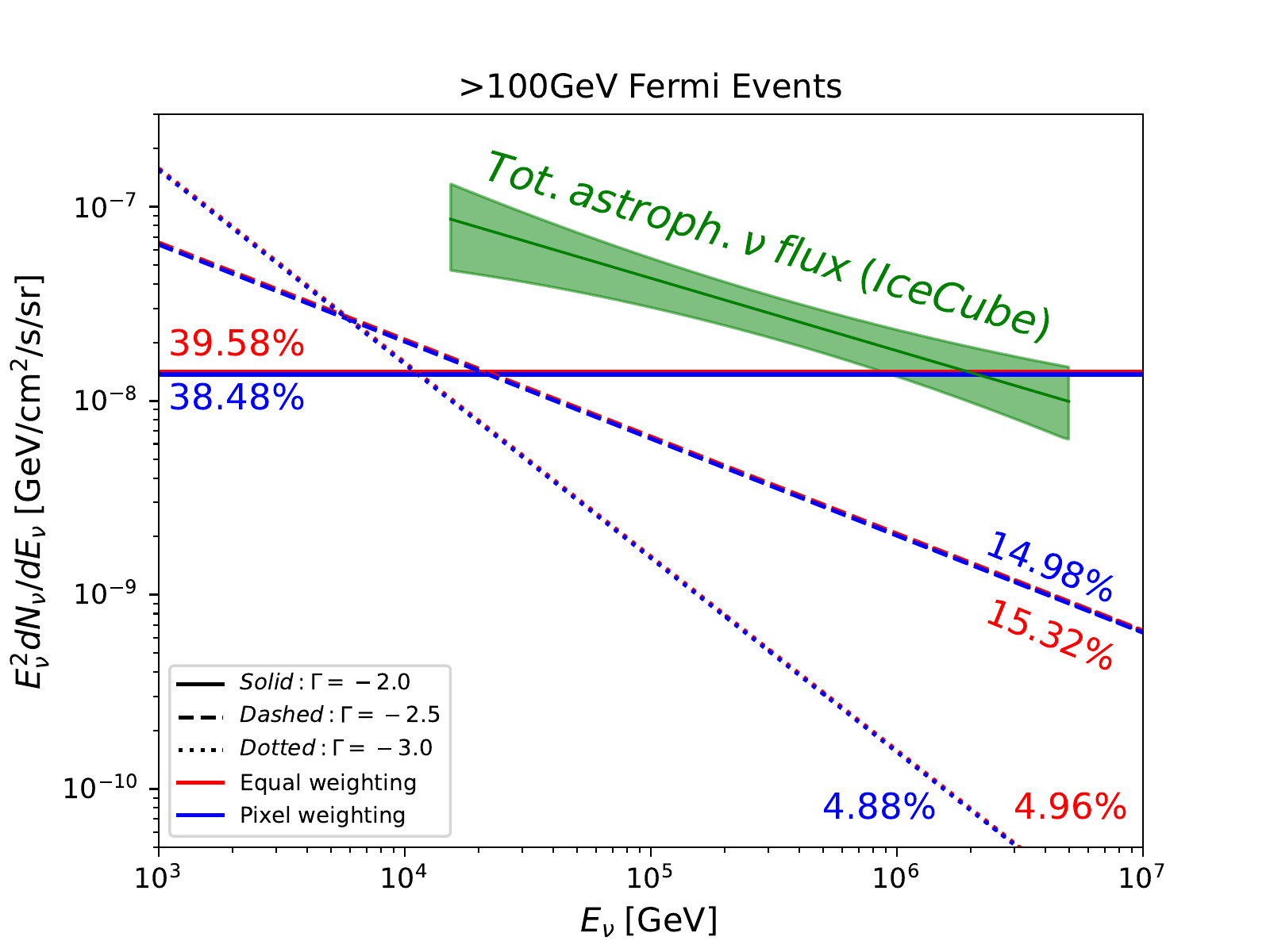}
\caption{We pixelate the HEEs into skymap first and then treat pixels as point sources to perform the likelihood analysis. {\it Left}: Change of the TS value as a function of the total neutrino flux for the HEE sample. {\it Right}: The 95\% confidence level upper limits on the total neutrino flux from the directions of the HEEs. {We notice that for the two weighting schemes nearly the same results are obtained in the right panel, which is in fact a coincidence. From the left panel, we can see that they have different likelihood profiles.}}
\label{fig_hee2}
\end{figure*}

In the left panel of Fig.~\ref{fig_100}, we present the result of the single-point-source analysis (i.e., the distribution of TS values in the directions of Fermi events with $E_\gamma>100\,{\rm GeV}$). Overall, the TS distribution is consistent with that predicted from Gaussian fluctuations of the background. We note that in the part of $\sqrt{\rm TS}>3$, there appears to be an excess in the number of large TS values compared to the uniform background expectation.
To quantify the significance of the excess, we adopt the formalism in Ref. \cite{Aartsen:2019fau} (see Eq.~(1) therein),
\begin{equation}
p_{\mathrm{bkg}}=\sum_{i=k}^{N} P_{\mathrm{binom}}\left(i \mid p_{k}, N\right)=\sum_{i=k}^{N}\left(\begin{array}{c} N \\ i \end{array}\right) p_{k}^{i}\left(1-p_{k}\right)^{N-i}.
\end{equation}
We find that the most significant $p_{\rm bkg}$ corresponds to $k=50$ (i.e., the excess is contributed by 50 most significant Fermi events) with a significance of $\sim4.8\sigma$. However, further examination reveals that 24 of the 50 events are {from the direction of the source PKS 1424+240}.
Relatively high significance ($\sim3.2\sigma$, pre-trial) of this source to be a potential neutrino source has already been reported \cite{Aartsen:2019fau}.
Removing the events {spatially associated} with this source from the analysis results in a significance of only $\sim1.7\sigma$ (pre-trial). Therefore, we do not find any new evidence of excess neutrino emission in this analysis.
In the right panel of Fig.~\ref{fig_100}, we perform a stacking analysis of all these $>100$ GeV events (treat each event as a source), which shows no evidence in support of excess neutrino emission.

Above we show that the overlaps of events from the same source will bias the results. We therefore alternatively using another way to compare the correlation between the HEEs and the high-energy neutrinos. We pixelate the 13335 HEEs into a HEALPix skymap (Fig.~\ref{fig:heemap}) with NPixels=49152 (i.e., NSide=64, corresponding to a pixel size of $\sim0.9^\circ$). Considering that the mean angular uncertainty of the IceCube events is also $\sim0.9^\circ$, we think this choice is fine enough to not loose too much sensitivity. Then, we treat pixels of the skymap with non-zero event numbers as point sources. We also use the number of events in each pixel as the weighting ($\omega_{j,{\rm model}}$). The results with this method are shown in Fig.~\ref{fig_hee2}. We still do not find a significant correlation between the HEE map and the IceCube data. {At most, only $\sim39\%$ of the IceCube diffuse neutrino flux can be come from the directions of these HEEs.}

\subsection{4FGL and 4LAC}
\label{sec_res_4fgl4lac}

\begin{figure}
\centering
\includegraphics[width=0.45\textwidth]{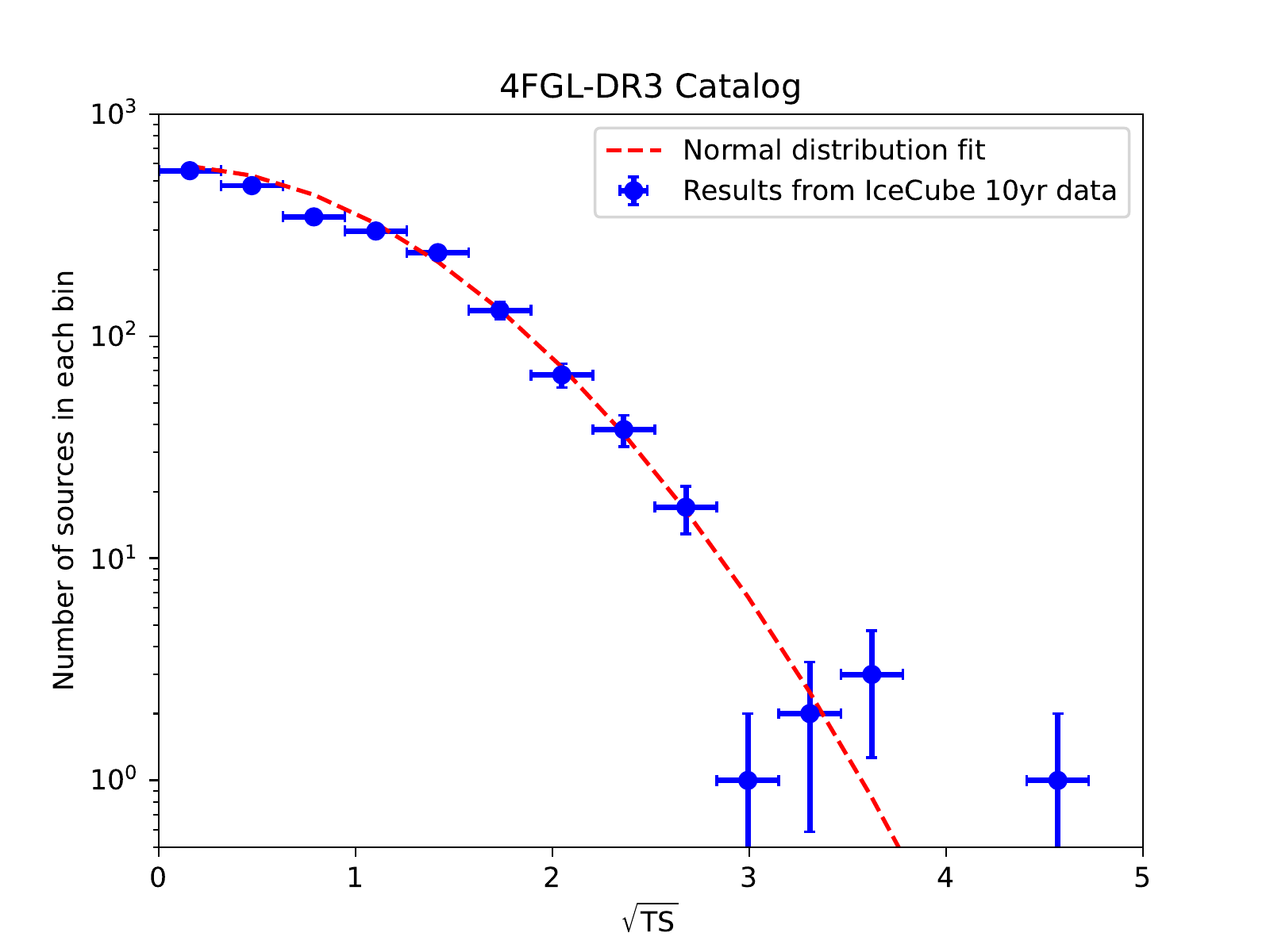}
\caption{Distribution of TS values for the sources in the 4FGL sample.
}
\label{fig_4fglts}
\end{figure}

\begin{table}
	\centering
	\caption{Number of overlap sources between samples.}
    \begin{tabular}{ccc}
		\hline
        \hline
        Catalog Name & $N_{\rm overlap}$ & $N_{\rm src}$\\
		\hline
        4FGL & \makecell{(2496 in 4LAC)\\(610 in 3FHL)} & 4689\\
        \hline
        4FGL Blazars & \makecell[c]{(2495 in 4LAC)\\(554 in 3FHL)} &  3399\\
        \hline
        4FGL BL Lacs & \makecell[c]{(1010 in 4LAC)\\(385 in 3FHL)} & 1354\\
        \hline
        4FGL FSRQs & \makecell[c]{(570 in 4LAC)\\(77 in 3FHL)} & 757\\
		\hline
        4LAC &\makecell[c]{(586 in 3FHL)} & 3128\\
        \hline
        4LAC Blazars &\makecell[c]{(555 in 3FHL)} & 3060\\
        \hline
        4LAC BL Lacs &\makecell[c]{(386 in 3FHL)} & 1236\\
        \hline
        4LAC FSRQs & \makecell[c]{(77 in 3FHL)} & 707\\
		\hline
        3FHL & $-$ & 1215\\
        \hline
        3FHL Blazar & $-$ &1078\\
        \hline
        3FHL BL Lacs & $-$ & 697\\
        \hline
        3FHL FSRQs & $-$ & 165\\
		\hline
		\hline
	\end{tabular}
\begin{tablenotes}
\item $^*$ The 4FGL refers to 4FGL-DR3 and 4LAC refers to 4LAC-DR2. \\
\item $^{**}$ Two sources in different catalogs are regarded as an overlap source if their spatial positions are consistent within an error circle of $0.1^\circ$.
\end{tablenotes}
\label{tab2}
\end{table}

Finally, we consider the 4FGL and 4LAC samples. Previously, there have been many works that searched for neutrino emissions from sources belonging to Fermi-LAT AGN catalogs (2LAC \cite{IceCube:2016qvd}, 3LAC \cite{Hooper:2018wyk}, 4LAC \cite{Smith:2020oac}). Compared to the 4LAC, the 4FGL catalog contains additional unassociated point sources. The majority of the unassociated sources at high latitudes are also AGN origins. 
{The overlaps between different samples tested in this work are summarized in Table~\ref{tab2}. As seen in the table, the  4FGL and 3FHL have only a small fraction of overlapping sources and they can be regarded as roughly independent samples. However, the 4FGL and 4LAC have more source overlap, so these two samples cannot be regarded as completely independent. However for completeness, we also present results based on both catalogs.}

Here we adopt the latest 4FGL-DR3~\cite{Fermi-LAT:2019yla,Fermi-LAT:2022byn} and 4LAC-DR2~\cite{Fermi-LAT:2019pir,Lott:2020wno} catalogs, as well as 10 years of IceCube data. The same as the 3FHL analysis (Section~\ref{sec_res_3fhl}), we exclude sources with $|b|<10^\circ$ and $|{\rm Dec}|>87^\circ$. 
The searches in the directions of individual 4FGL sources give results shown in Fig.~\ref{fig_4fglts}. Like the above 3FHL and HEE searches, the distribution is consistent with Gaussian fluctuations, with no sign of statistically significant neutrino emission. {We note there are 7 sources with TS$>$9, they are (the numbers in parentheses are TS values): NGC 1068 (22.3), J1210.3+3928 (13.6), J0006.40+0135 (12.9), J0940.3-7610 (12.7), J1543.0+6130 (10.3), J0118.7-0848 (10.0), PKS 1424+240 (9.7). The 4FGL sample contains a total of 4689 sources, statistically $\sim$12 ($\sim$2) sources are expected to have TS values greater than 9.0 (12.0). Therefore, our results here are consistent with the expectation from statistical fluctuations.} The 4LAC sample is roughly a subset of the 4FGL sample, so the individual-source search for this sample the same gives null results. 

In the stacking analyses of neutrino emission from 4FGL or 4LAC sources, for any combination of spectral indices and weighting schemes (see Section~\ref{sec_ana}), we are not able to identify any evidence of neutrino emission (see the Appendix for related figures). The sources in the 4FGL and 4LAC contribute at most 42.78\% and 57.69\% of the diffuse neutrino flux, respectively. Finally, we repeat our analysis for 3 subsets (Blazar, BL Lac, FSRQ). 
These analyses, again, show no evidence of neutrino emission. The relevant results are presented in the Appendix. 

\section{Summary and Conclusions}
High-energy astrophysical neutrinos provide a crucial window to study the Universe. However, at present, we know little about the origins of these neutrinos. Great efforts have been paid to find out the sources of the astrophysical neutrinos. Encouraging results that have been reported include the discovery of tentative neutrino excesses in the directions of TXS 0506+056 and NGC 1068. However, this is just the tip of the iceberg. These two sources, even if robustly confirmed as high energy neutrino emitters, contribute a very small fraction of the total high-energy neutrinos observed by the IceCube. Most of the diffuse astrophysical neutrinos remain unexplained.

In this paper, we study the correlations between IceCube high-energy neutrinos and gamma-ray observations by the Fermi-LAT telescope.
We use the muon-track neutrino data of ten years recently released by the IceCube collaboration and search for excess neutrino emission for various Fermi-LAT samples. The gamma-ray samples considered in the work include: 
3FHL, HEE, 4FGL, 4LAC and several subsets of these samples.
For each sample, both a single-source analysis and a stacking analysis are performed.
Compared to previous works of the same type, one of the major improvements in our analysis is that we utilize a larger (10 years) data set of IceCube neutrino observations.

No neutrino point source with large significance is found in our analysis. The stacking analysis also shows no statistically significant correlation between IceCube events and any of considered Fermi-LAT samples.
From the null searching results, we place upper limits on their contributions to the IceCube diffuse neutrino flux. We conclude that in the case of $\Gamma=-2.5$ each of these source populations accouts for no more than $\sim$2.5\%-36\% of IceCube diffuse neutrino flux between 16 TeV and 2.6 PeV. 
Since we are using the latest 10 years of IceCube neutrino data, we obtain upper limits a factor of 2 stronger than those based on 3 years of data. We also note that the analyses of BL Lacs usually give less restrictive constraints among all the samples we have considered; together with the tentative excess in the direction of TXS 0506+056, this suggests BL Lacs may contribute non-negligible fraction of the astrophysical flux observed by IceCube. This possibility could be tested with future neutrino telescopes such as IceCube-Gen2 \cite{IceCube-Gen2:2020qha,Clark:2021fkg}. 

\begin{acknowledgments}
We thank for the valuable suggestion from the anonymous referee. We thank Bei Zhou for helpful discussions. This work is supported by the Guangxi Science Foundation (grant No. 2019AC20334) and Bagui Young Scholars Program (LHJ).
\end{acknowledgments}

\bibliographystyle{apsrev4-1-lyf}
\bibliography{refer.bib}

\widetext
\newpage

\appendix

\section{Figures for all samples}

\begin{figure*}[b]
\centering
\includegraphics[width=0.33\textwidth]{3fhl_tsdnde.pdf}
\includegraphics[width=0.33\textwidth]{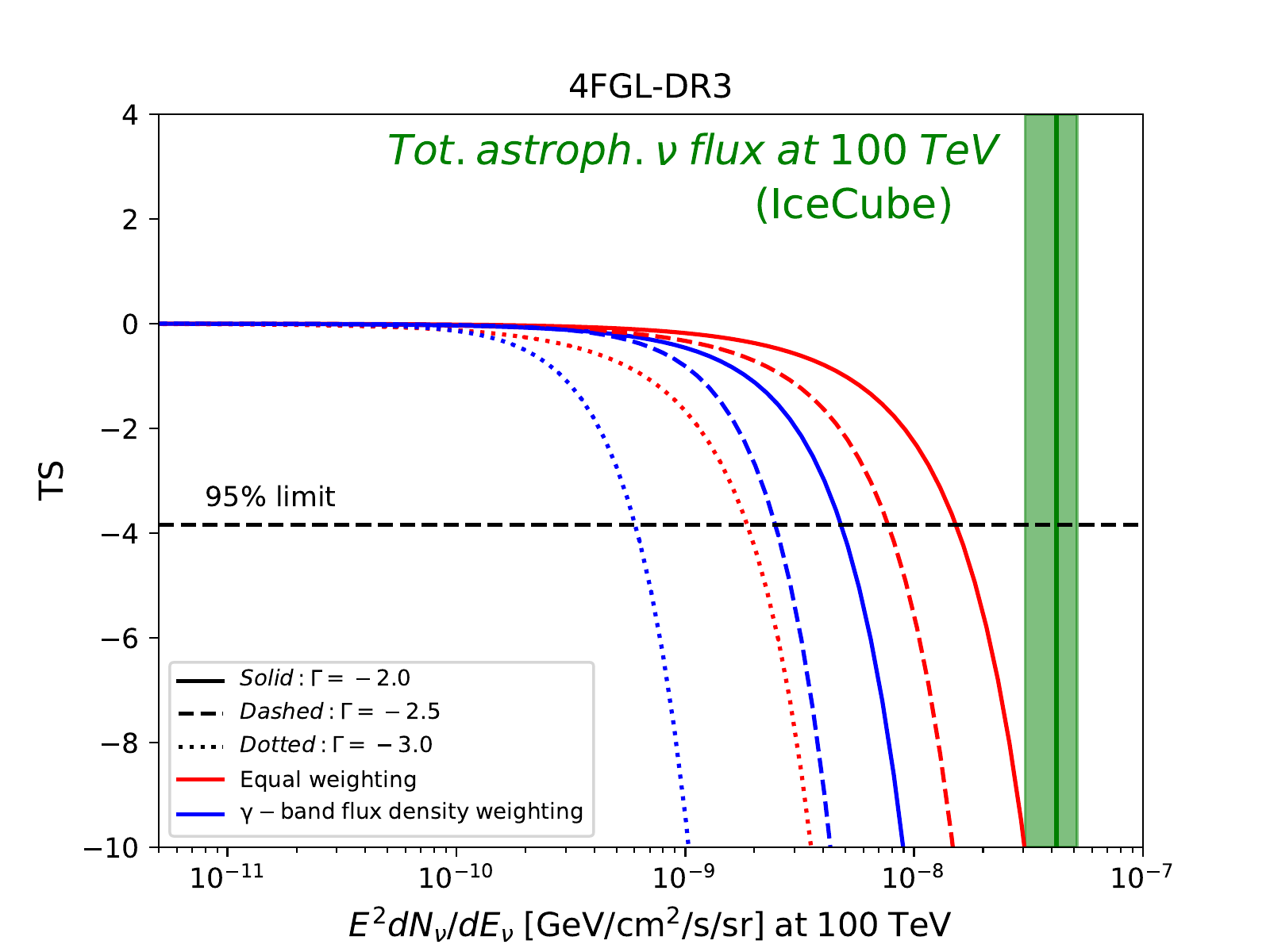}
\includegraphics[width=0.33\textwidth]{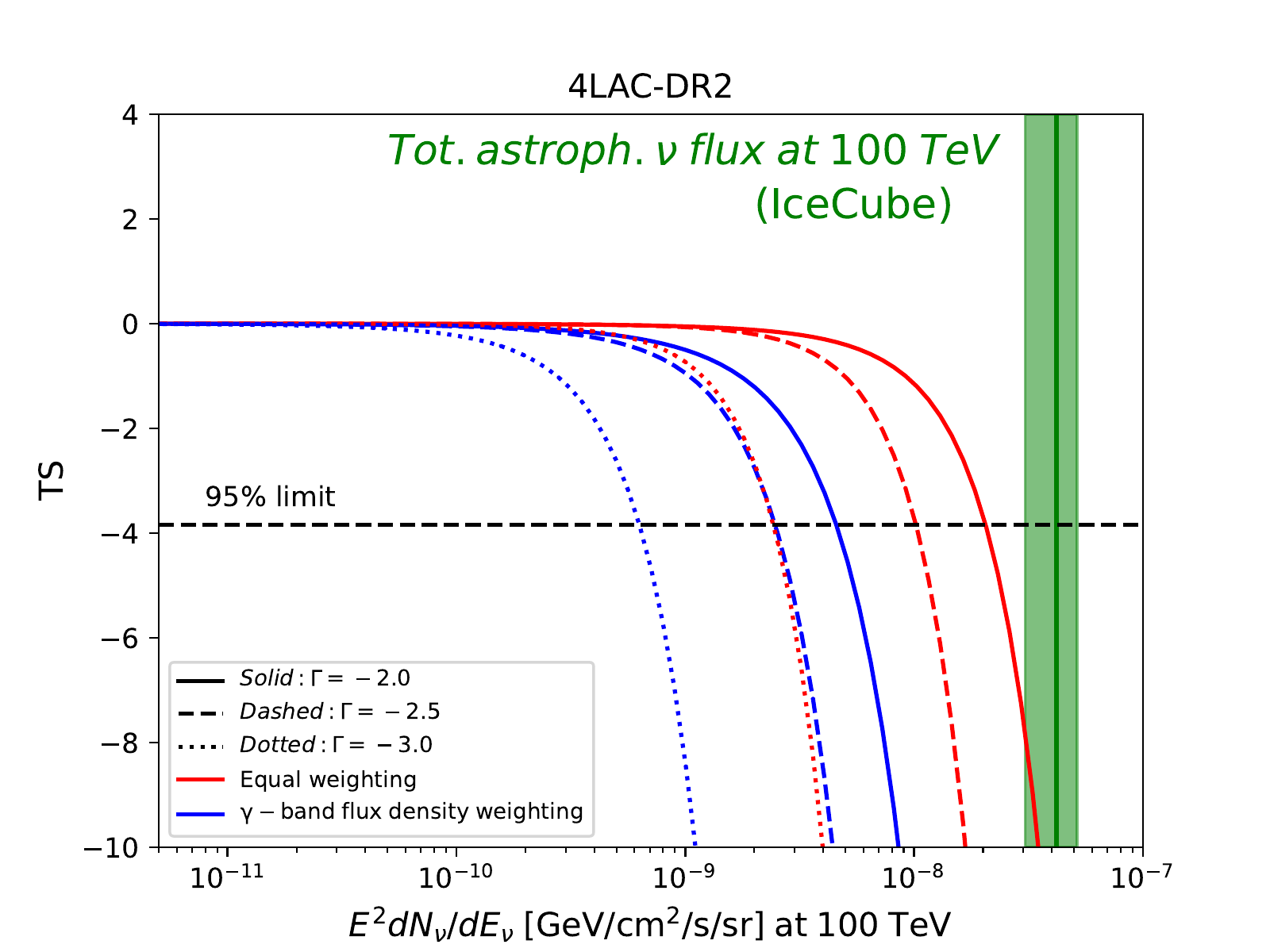}
\includegraphics[width=0.33\textwidth]{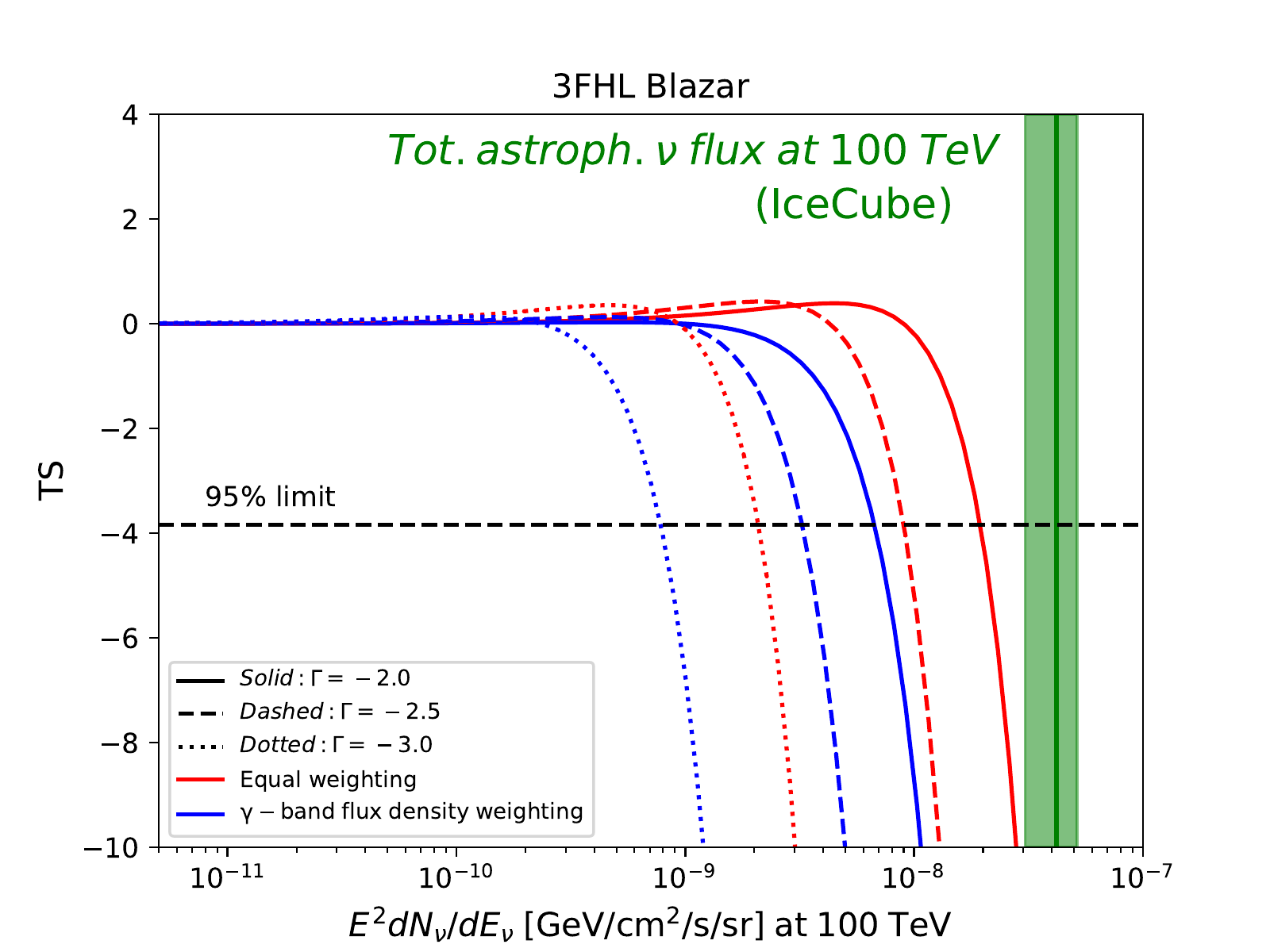}
\includegraphics[width=0.33\textwidth]{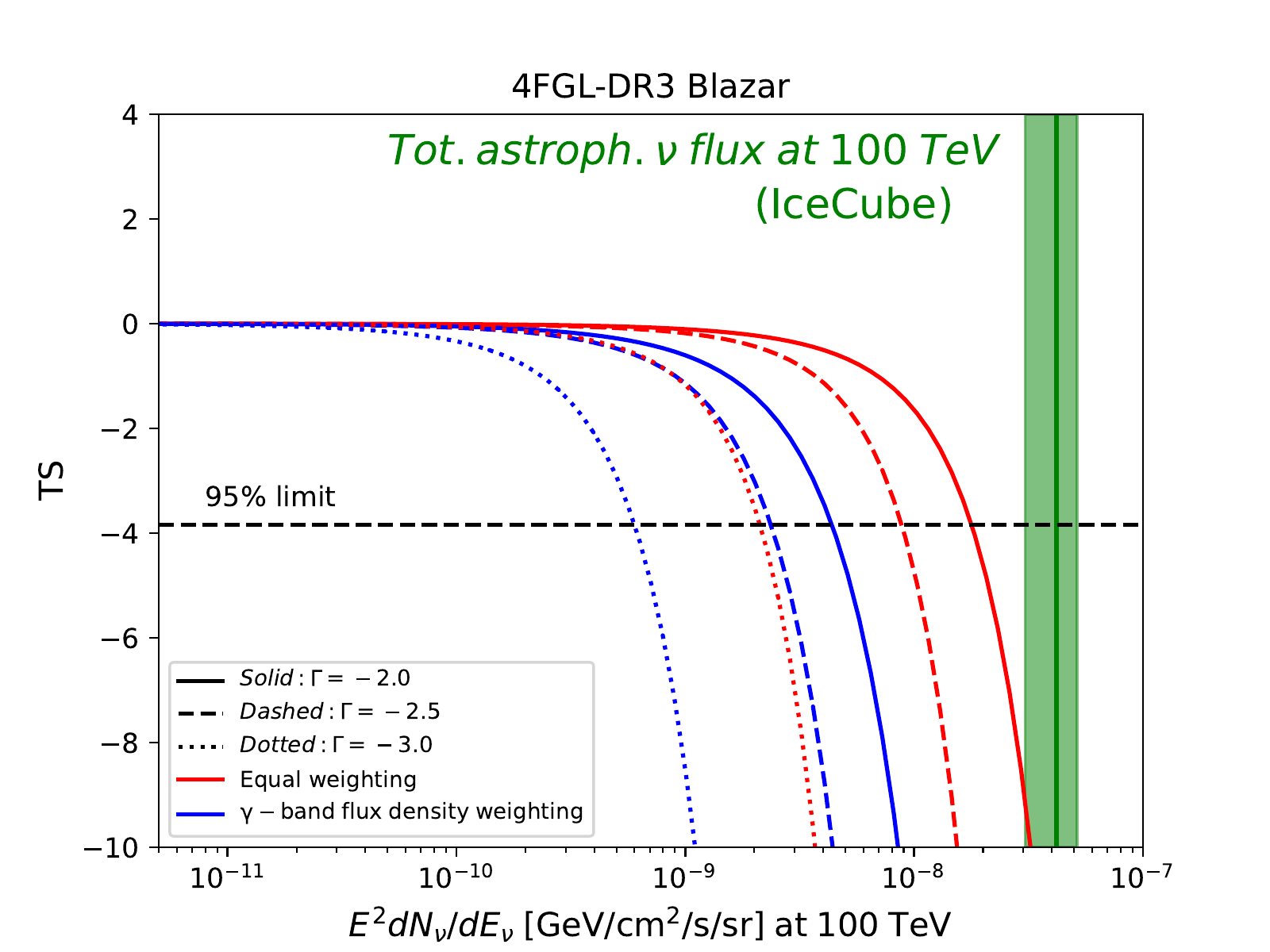}
\includegraphics[width=0.33\textwidth]{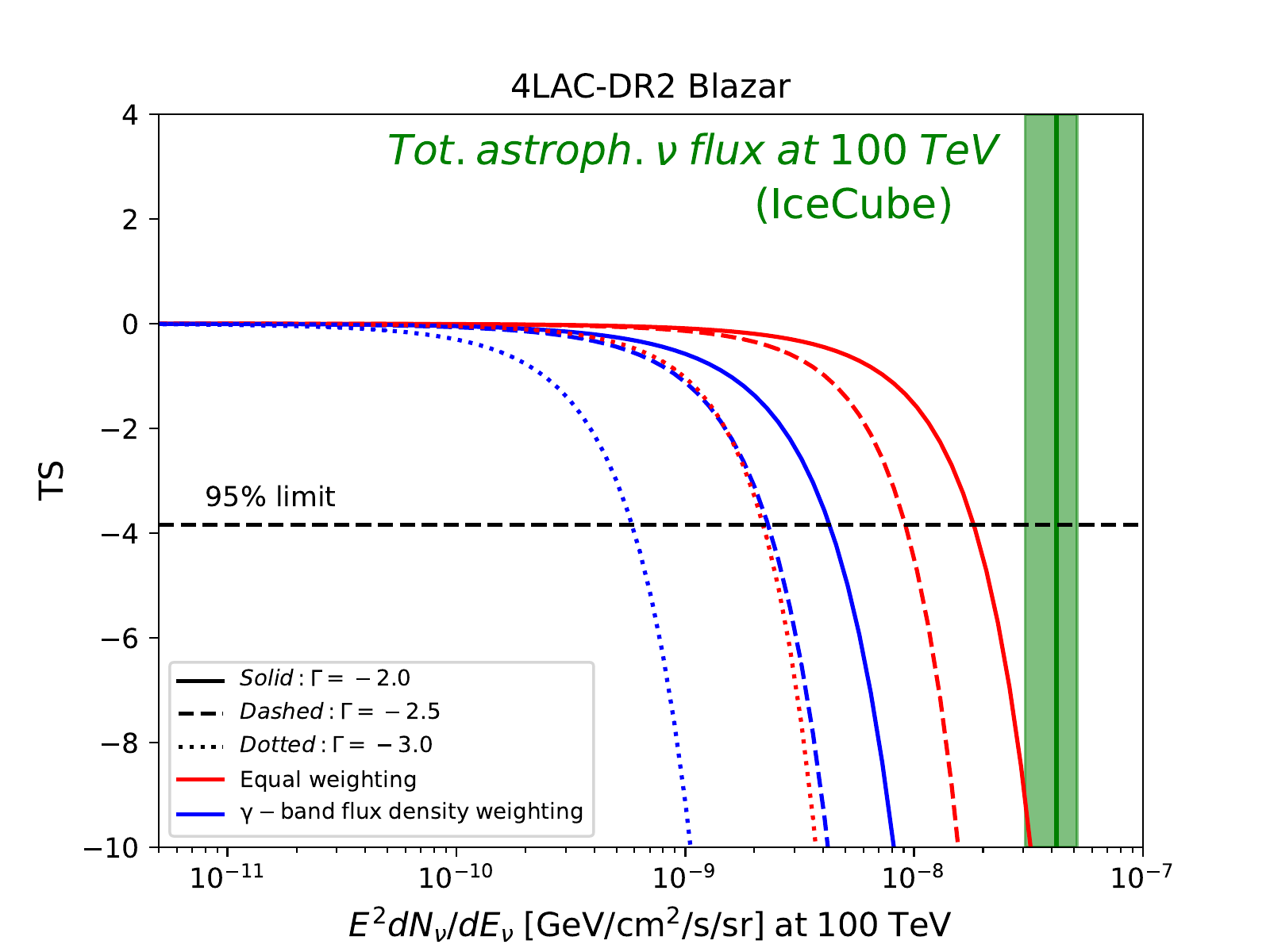}
\includegraphics[width=0.33\textwidth]{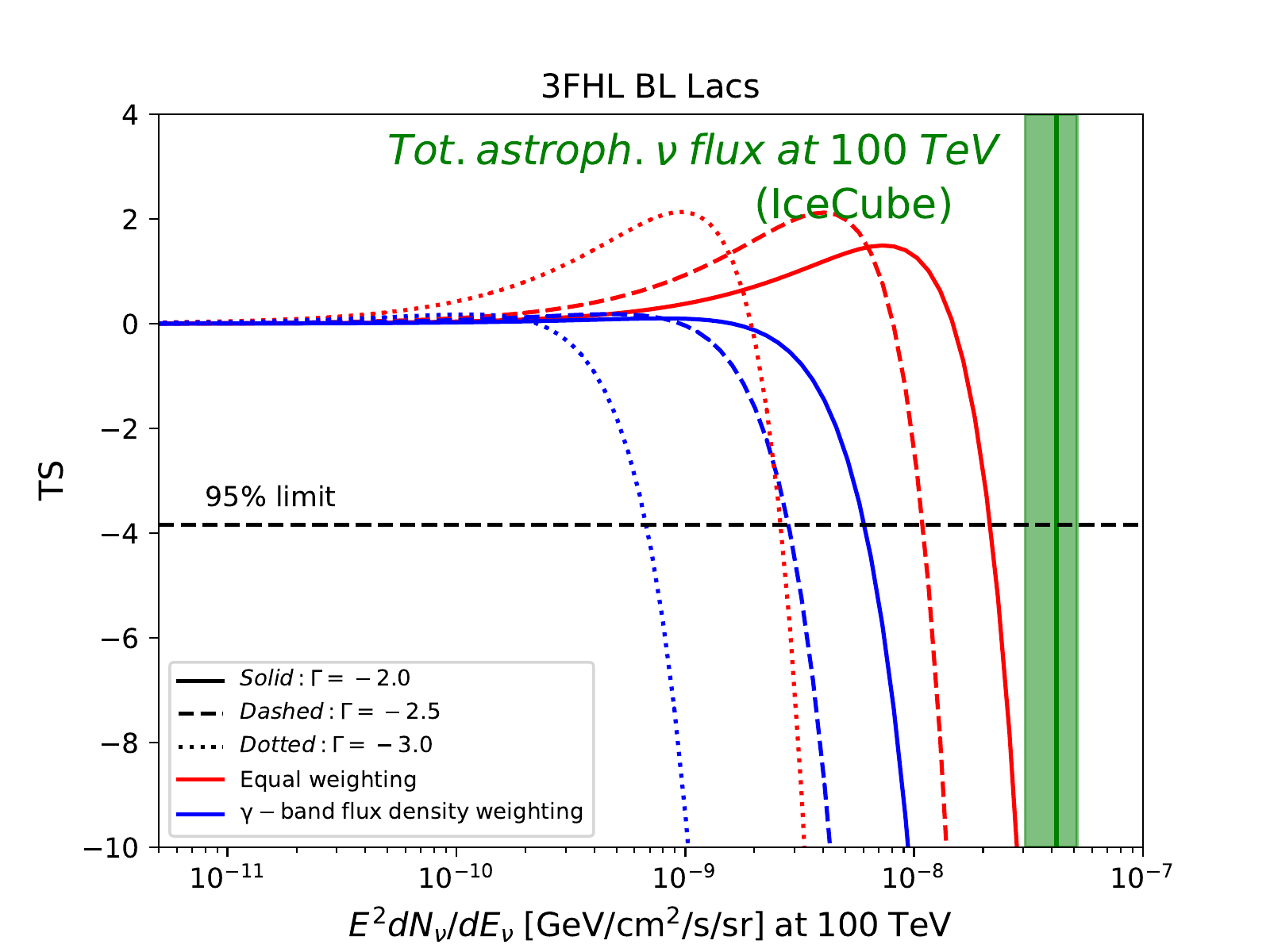}
\includegraphics[width=0.33\textwidth]{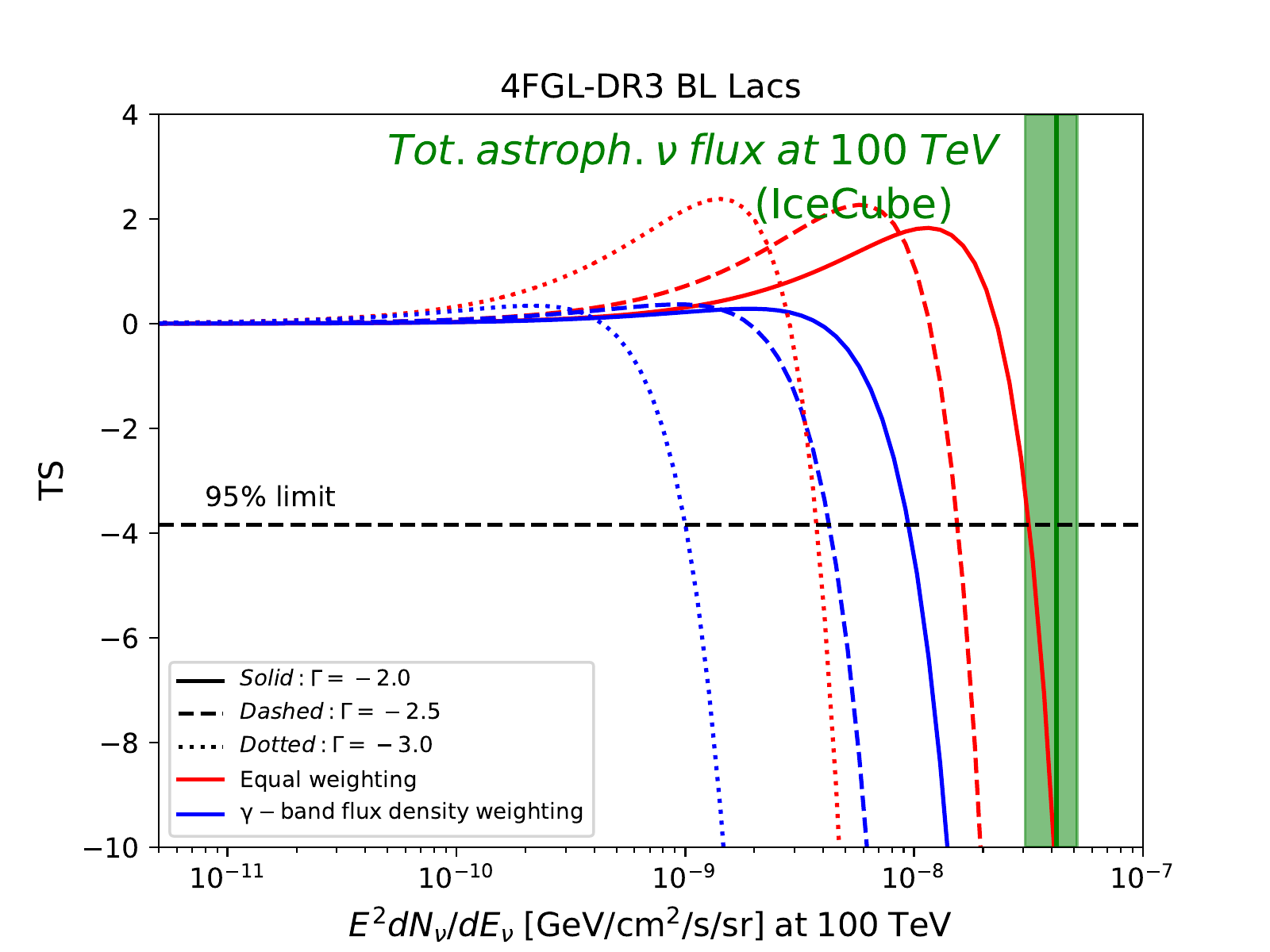}
\includegraphics[width=0.33\textwidth]{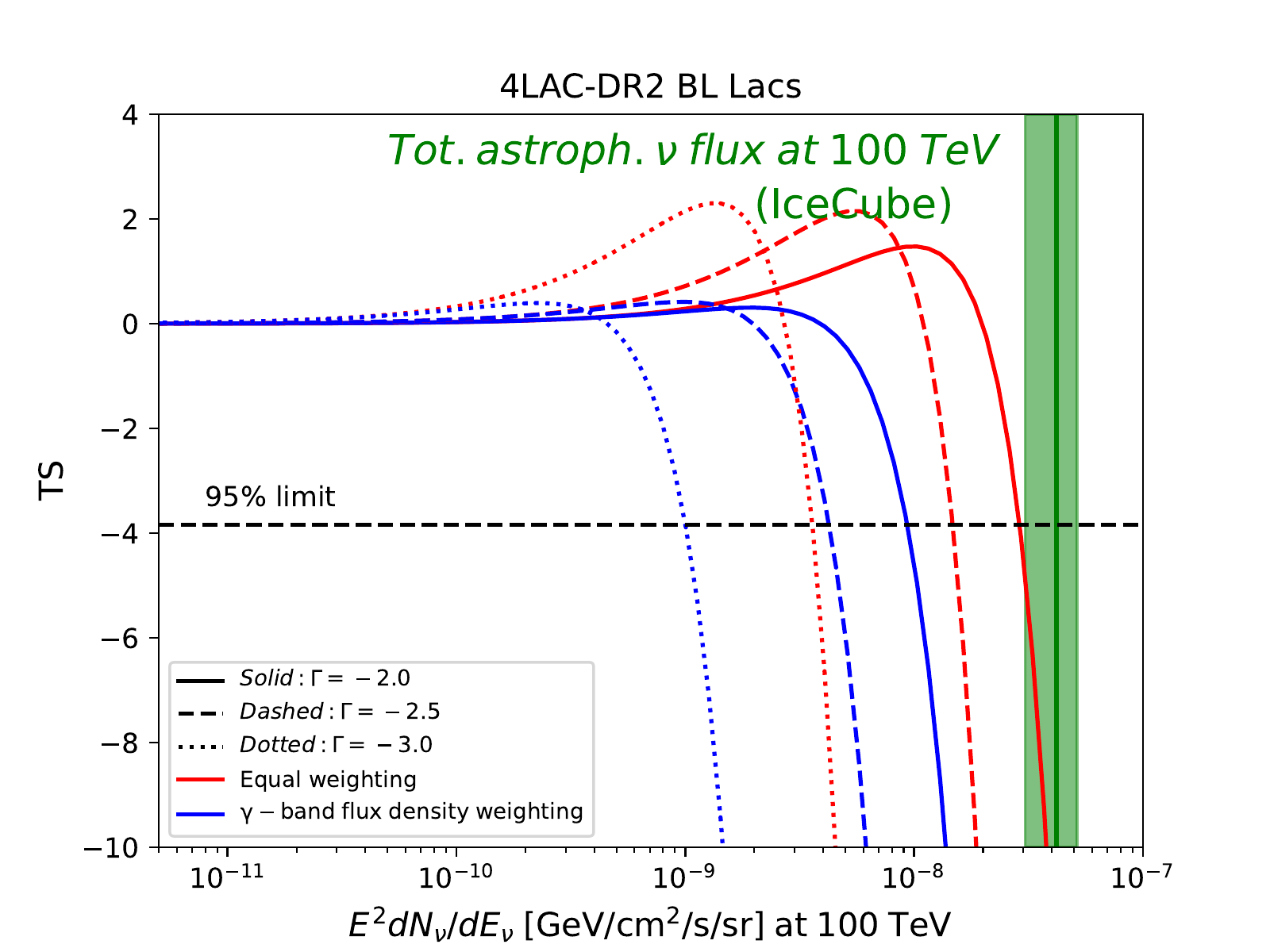}
\includegraphics[width=0.33\textwidth]{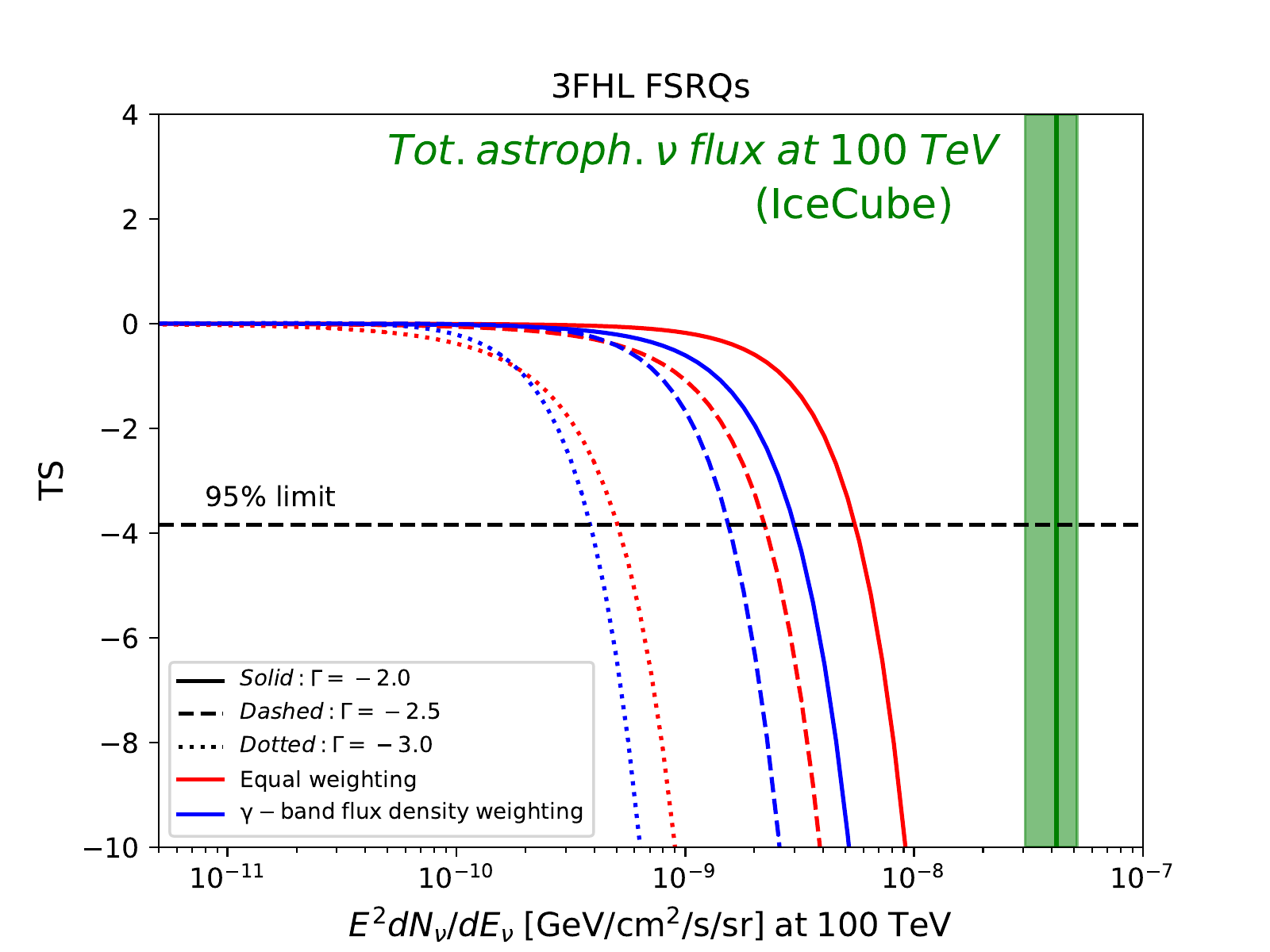}
\includegraphics[width=0.33\textwidth]{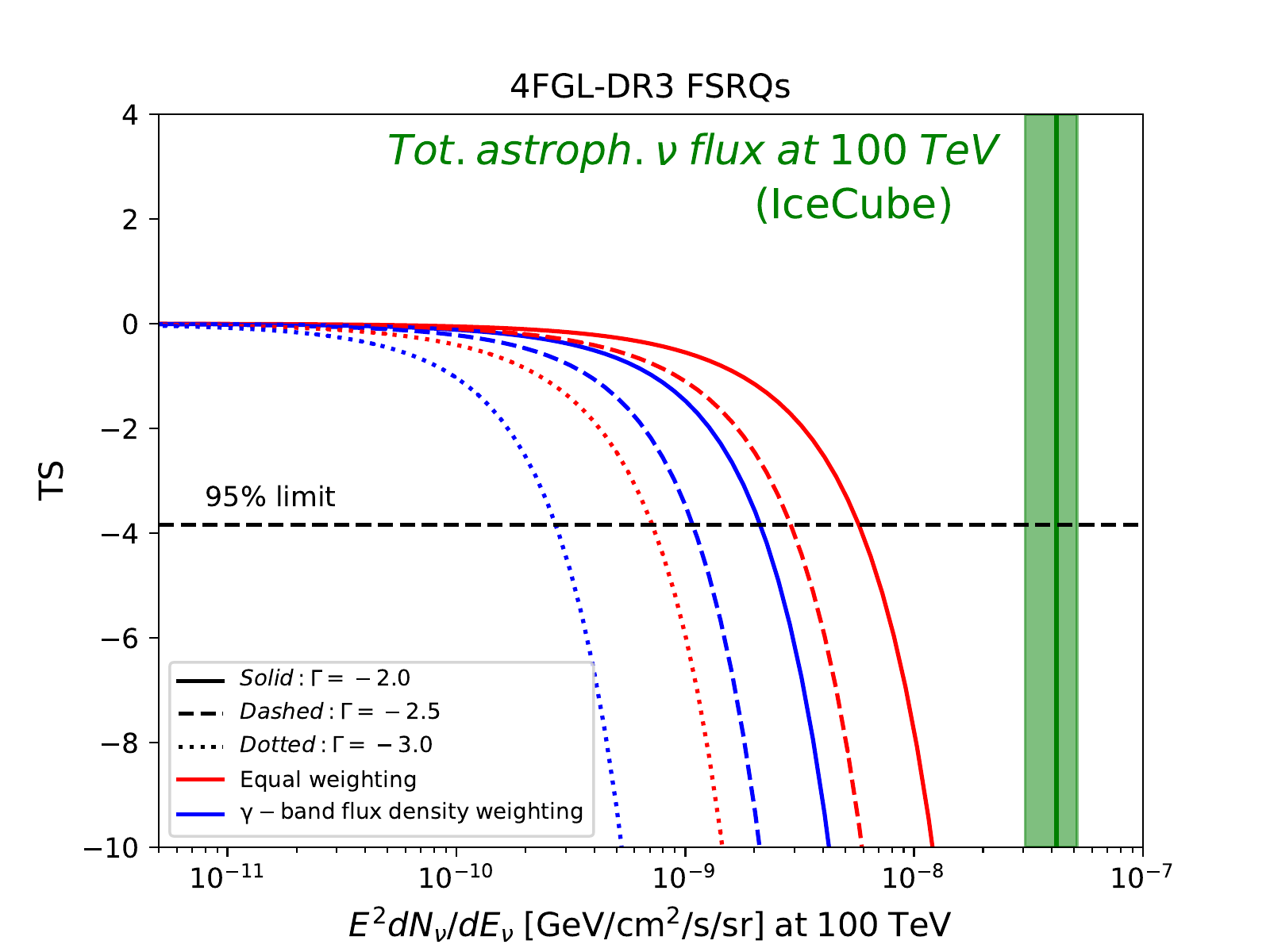}
\includegraphics[width=0.33\textwidth]{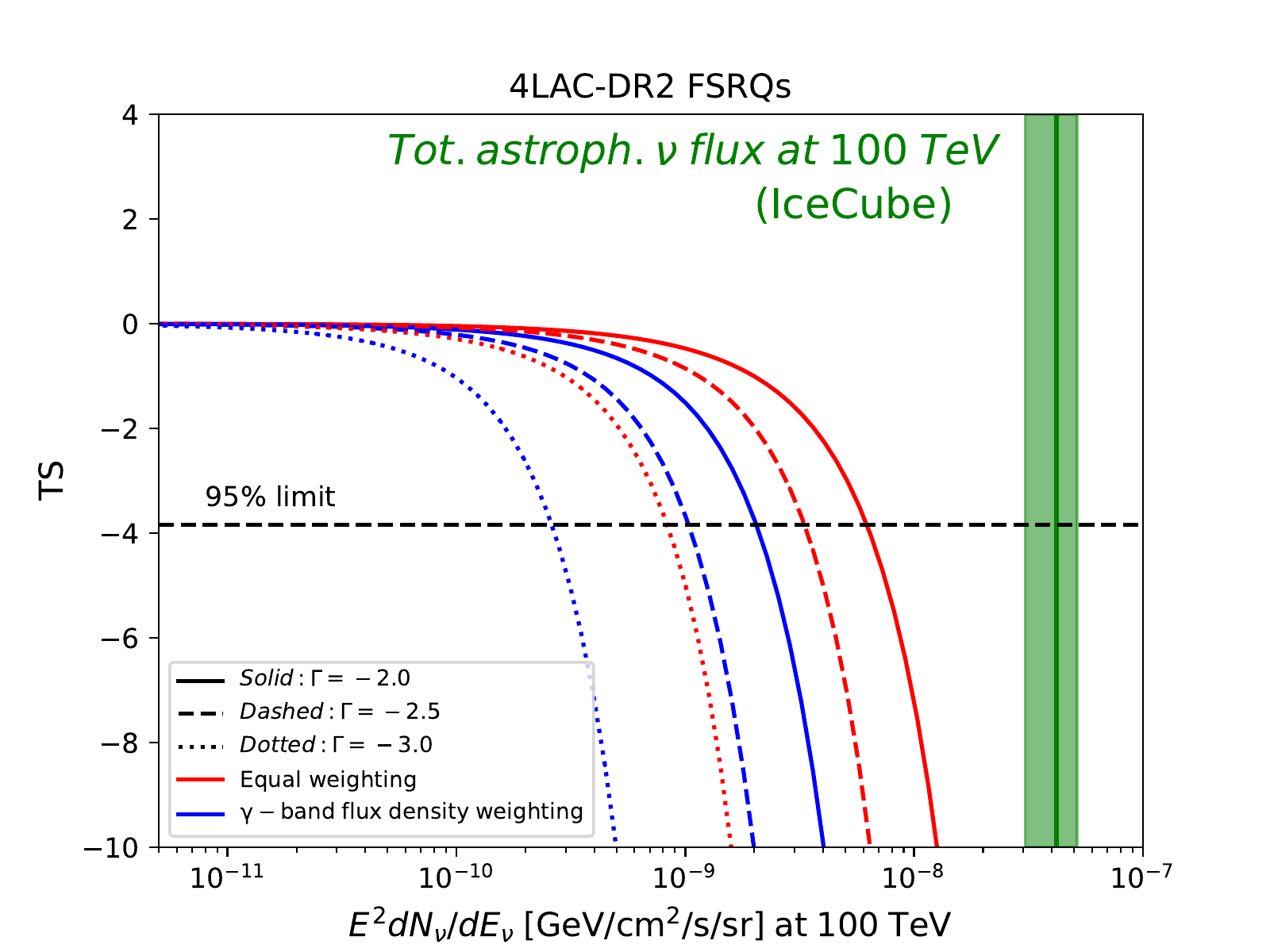}
\caption{The change of the TS value as a function of the total neutrino flux from the sources in different samples. The left, middle and right columns are for the 3FHL, 4FGL and 4LAC samples, respectively. The first row is for all sources in each catalog, while from the second to the fourth rows, the plots are for the subsets of blazars, BL Lacs and FSRQs, respectively. For each source sample, we consider three different choices of the neutrino spectral index (solid, dashed and dotted lines) and two weighting schemes (red and blue lines). The vertical green band in each plot is the all-sky diffuse astrophysical neutrino flux measured by IceCube.}
\end{figure*}

\begin{figure*}[h]
\centering
\includegraphics[width=0.33\textwidth]{3fhl_ednde.pdf}
\includegraphics[width=0.33\textwidth]{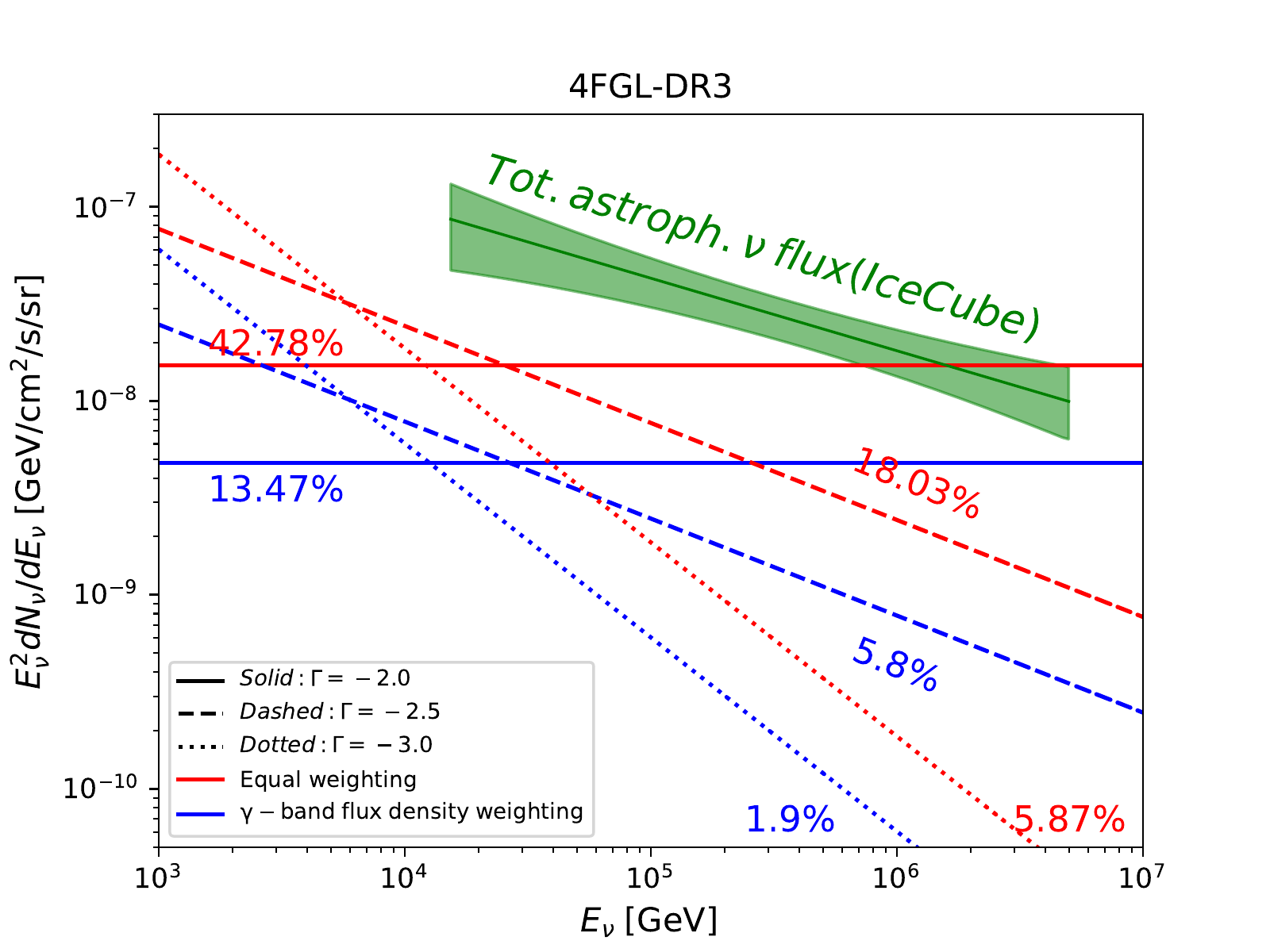}
\includegraphics[width=0.33\textwidth]{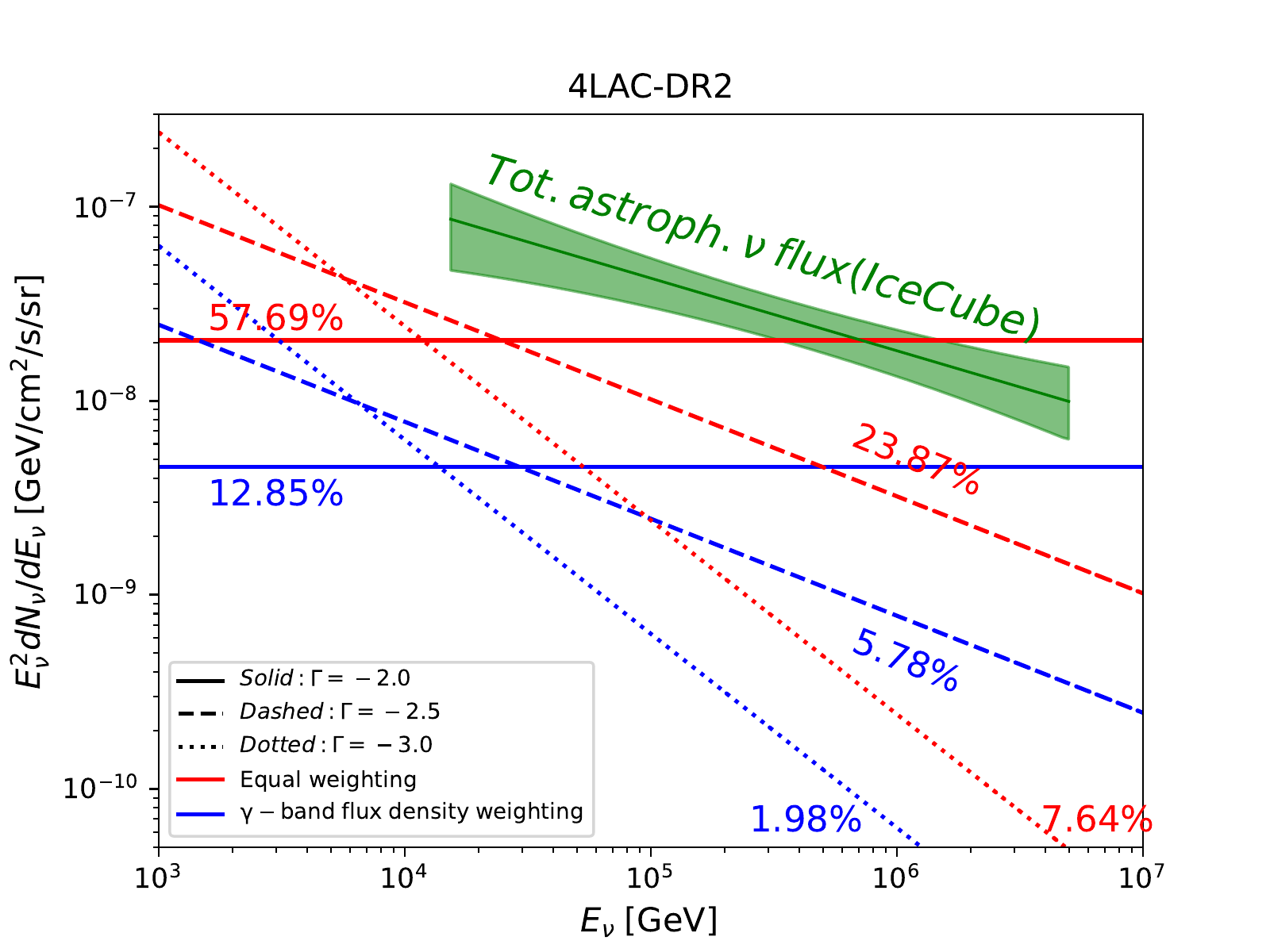}
\includegraphics[width=0.33\textwidth]{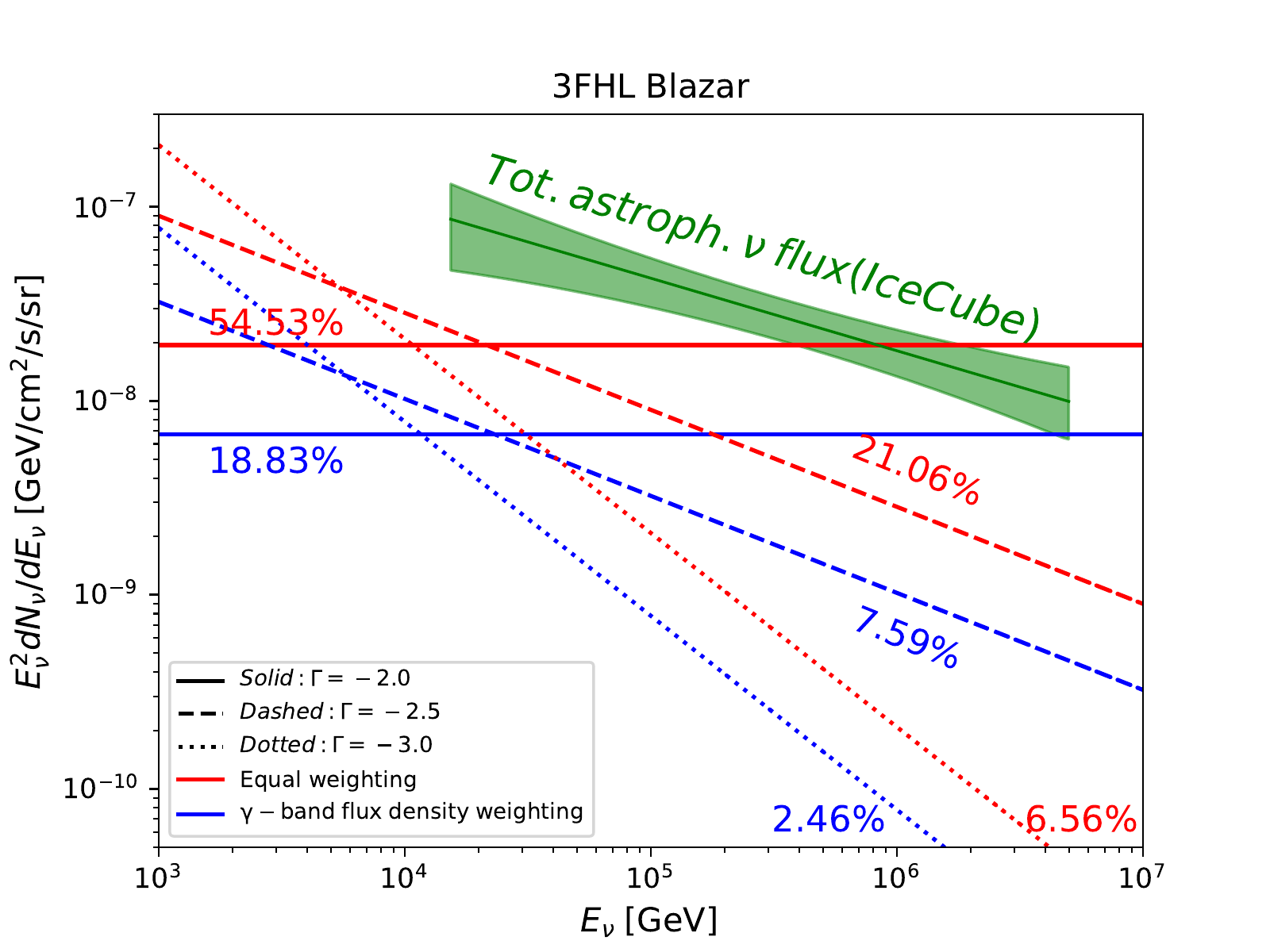}
\includegraphics[width=0.33\textwidth]{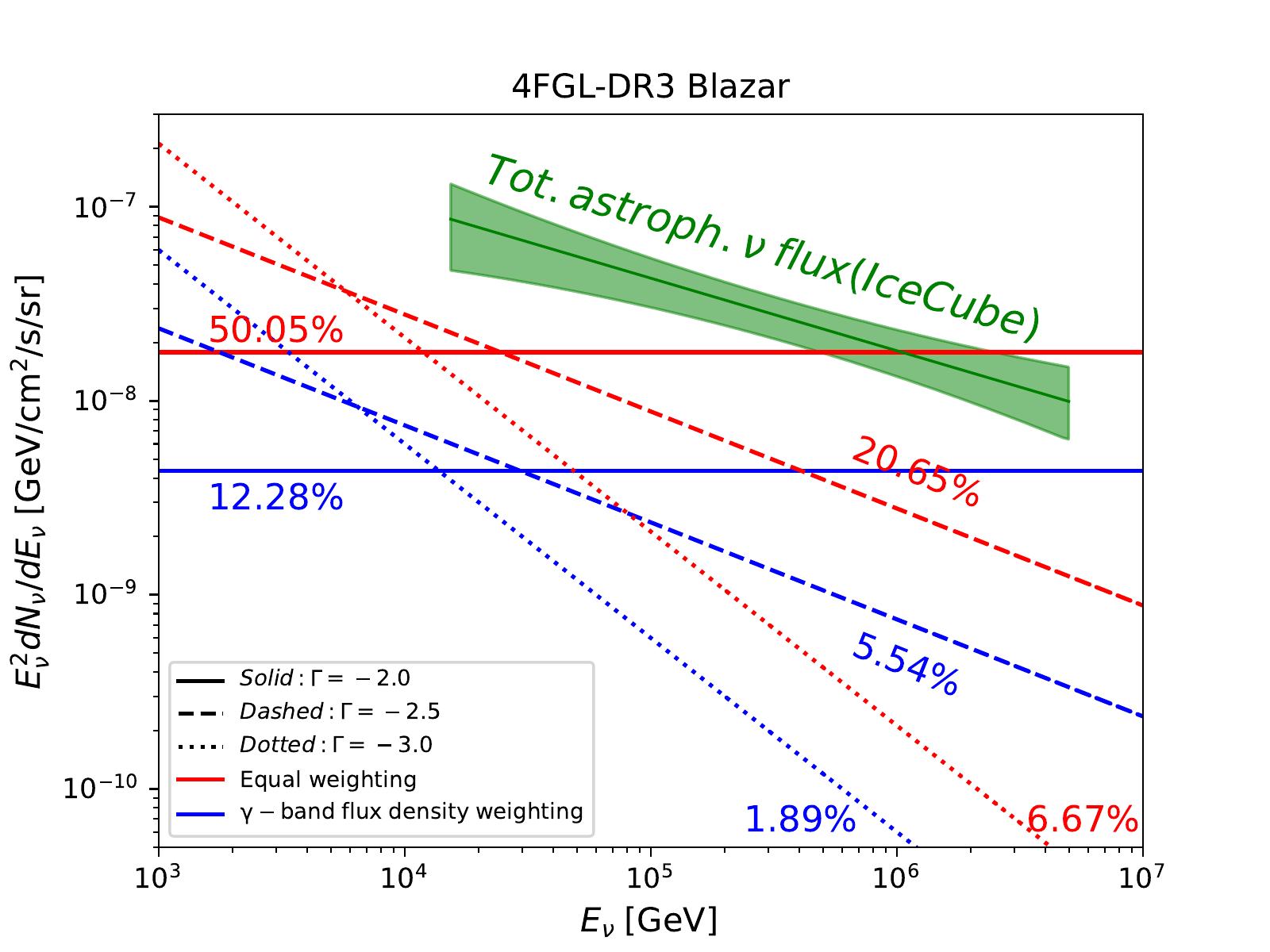}
\includegraphics[width=0.33\textwidth]{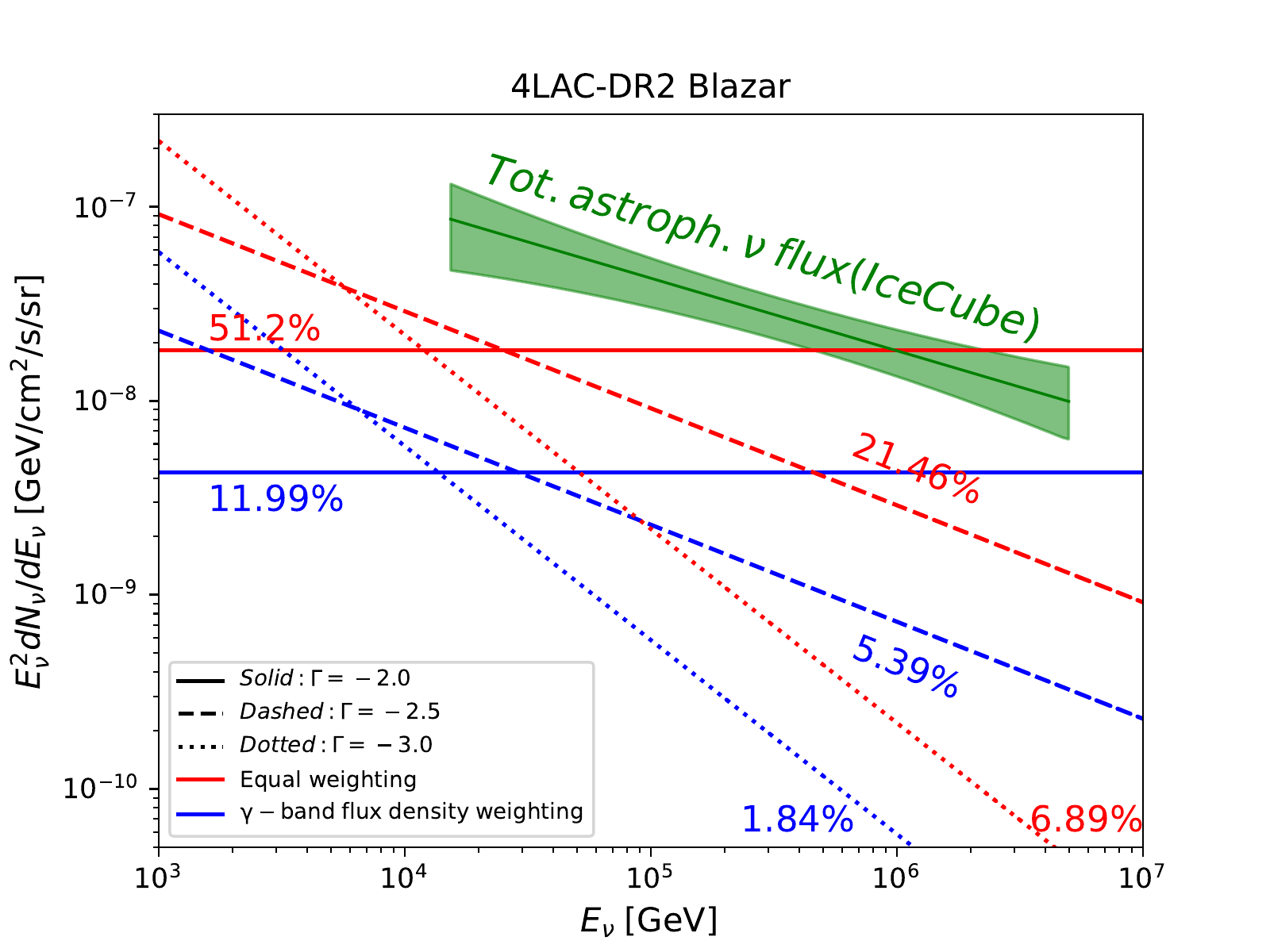}
\includegraphics[width=0.33\textwidth]{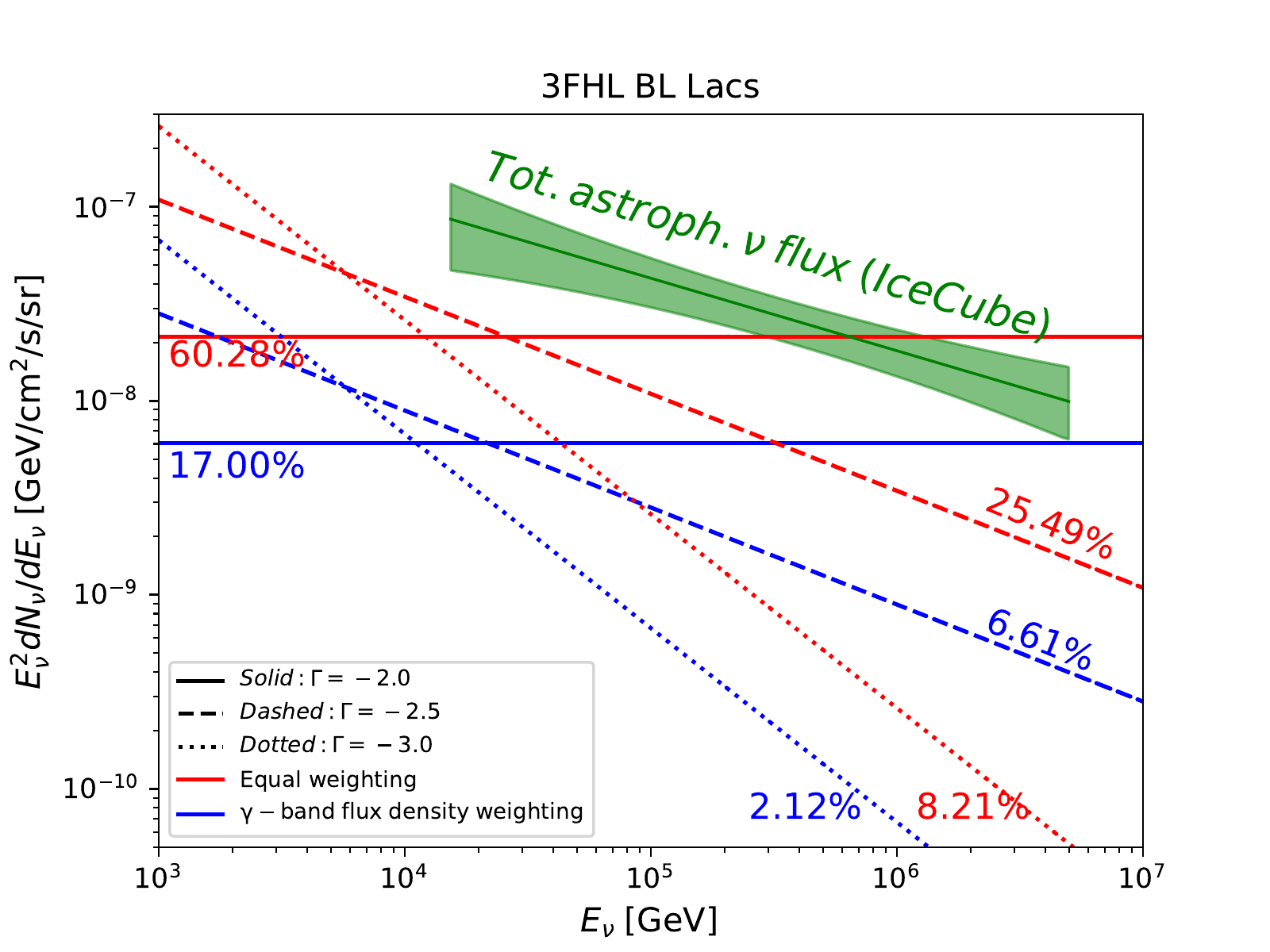}
\includegraphics[width=0.33\textwidth]{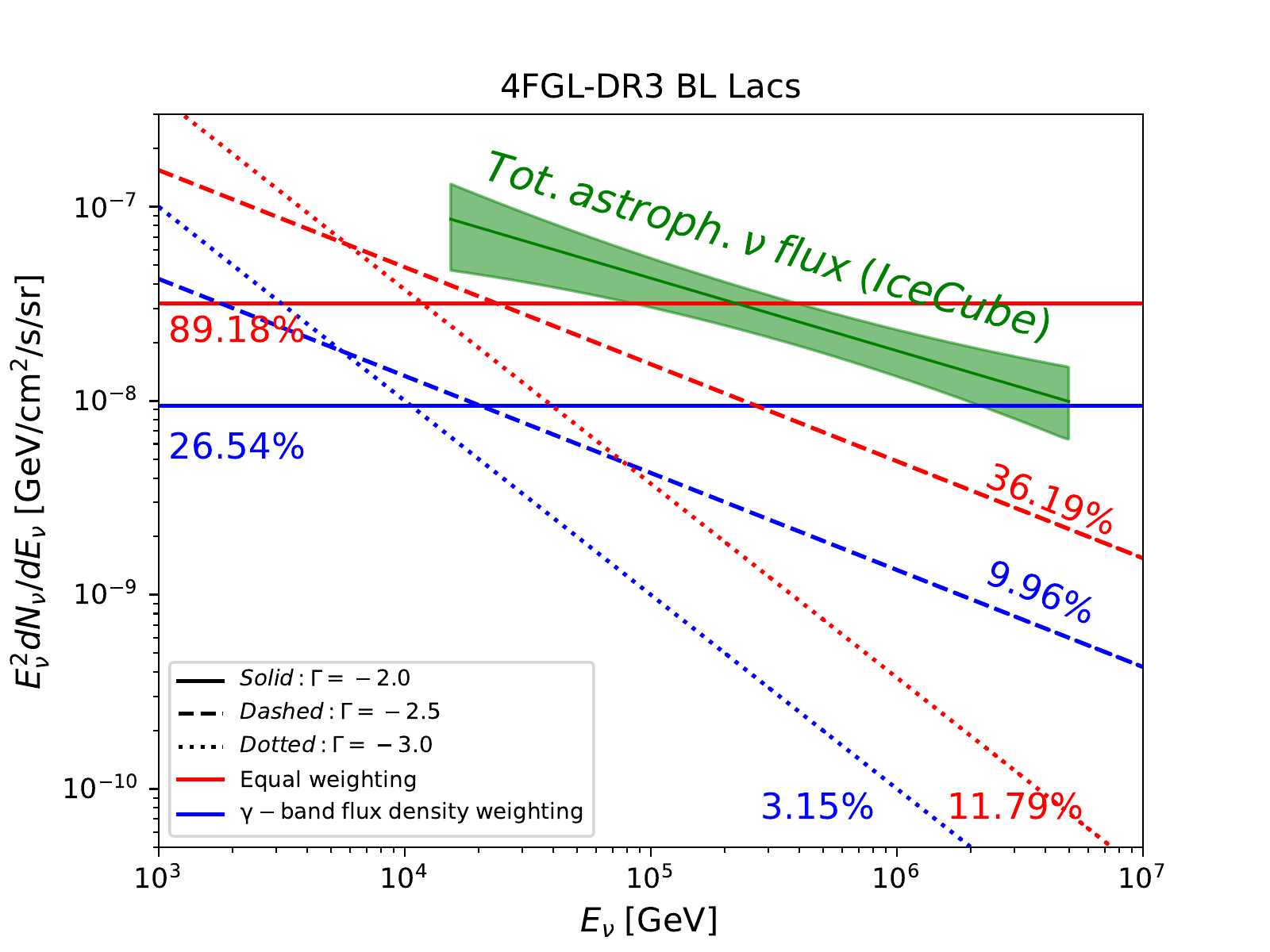}
\includegraphics[width=0.33\textwidth]{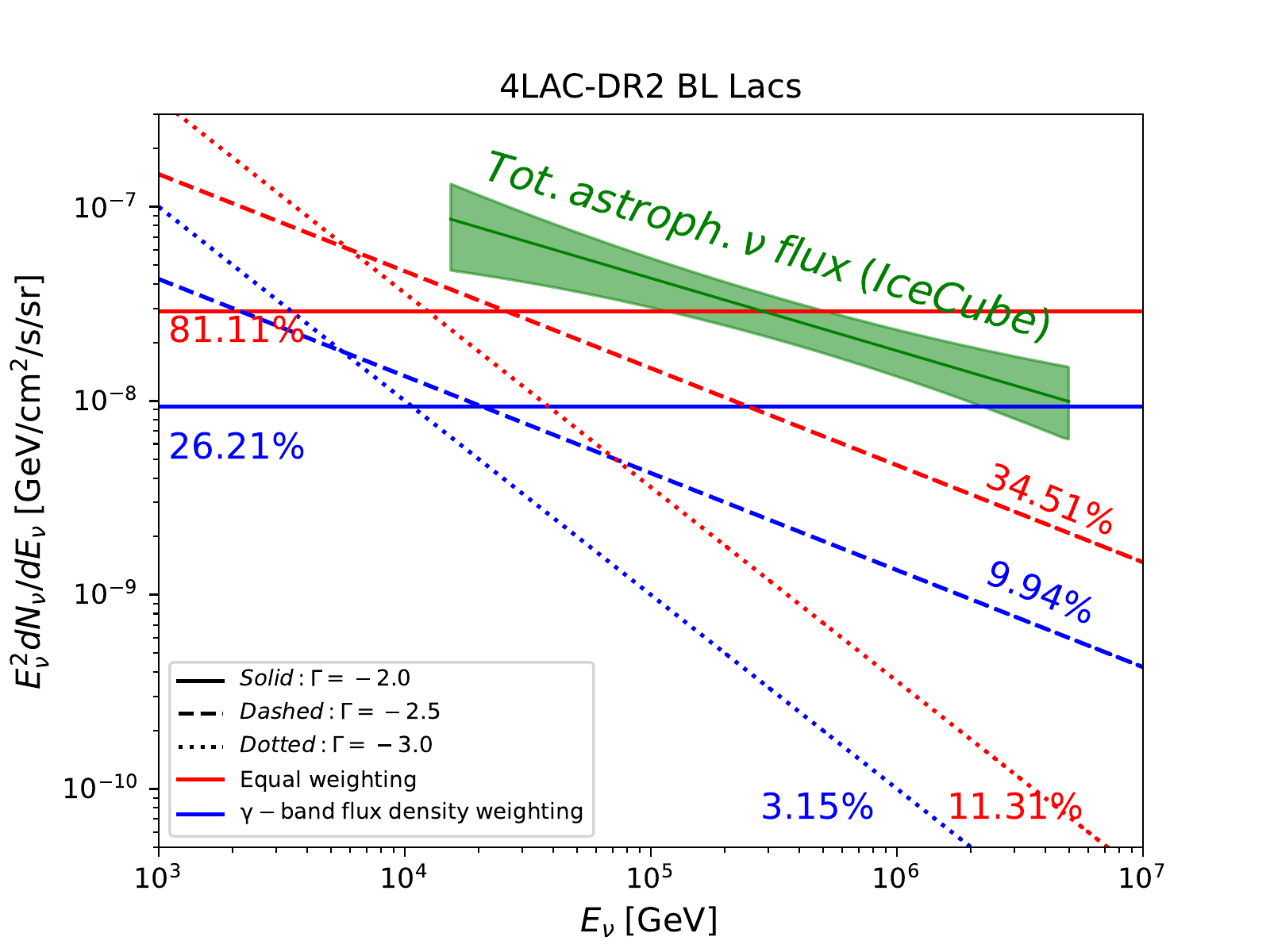}
\includegraphics[width=0.33\textwidth]{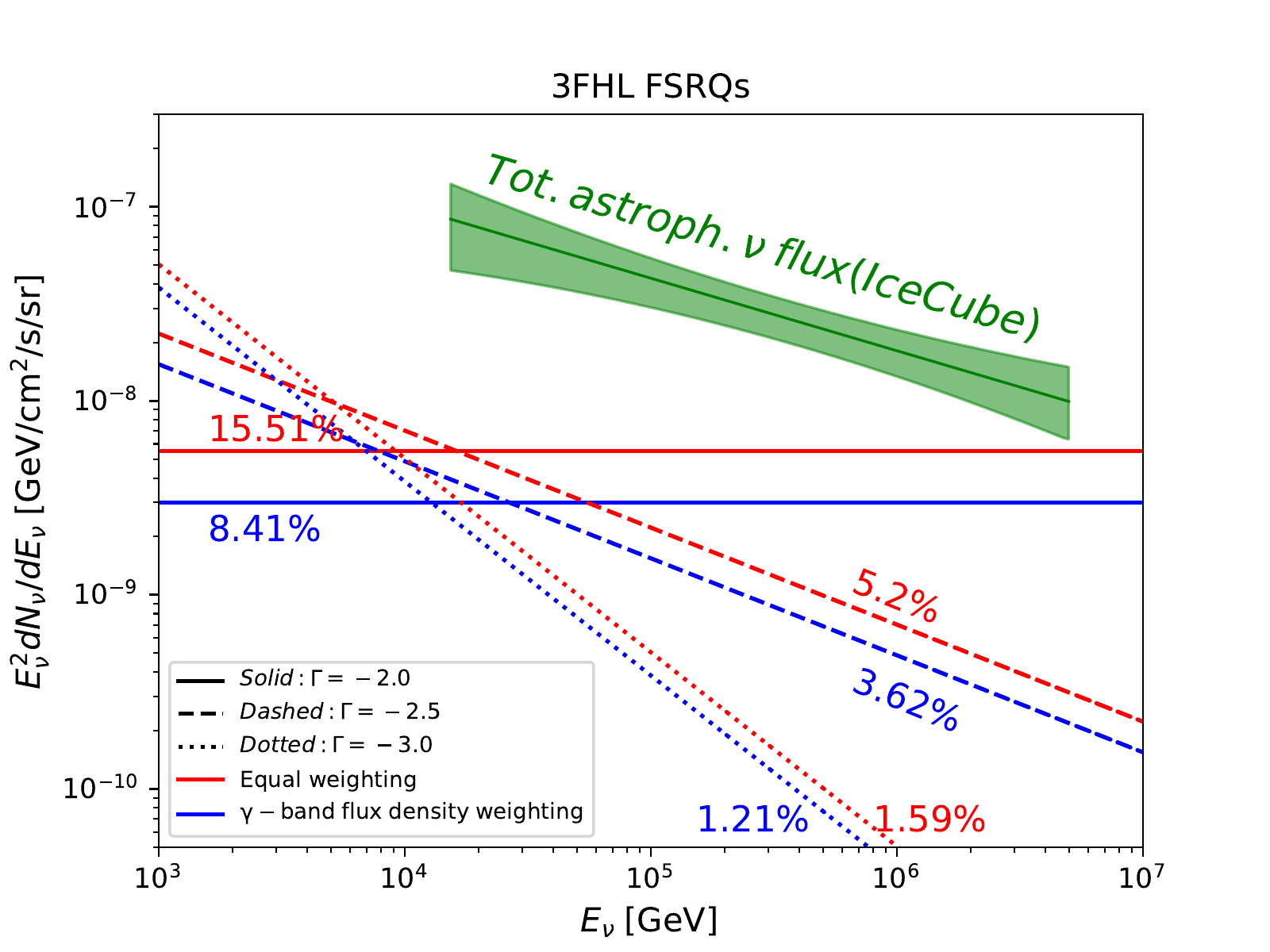}
\includegraphics[width=0.33\textwidth]{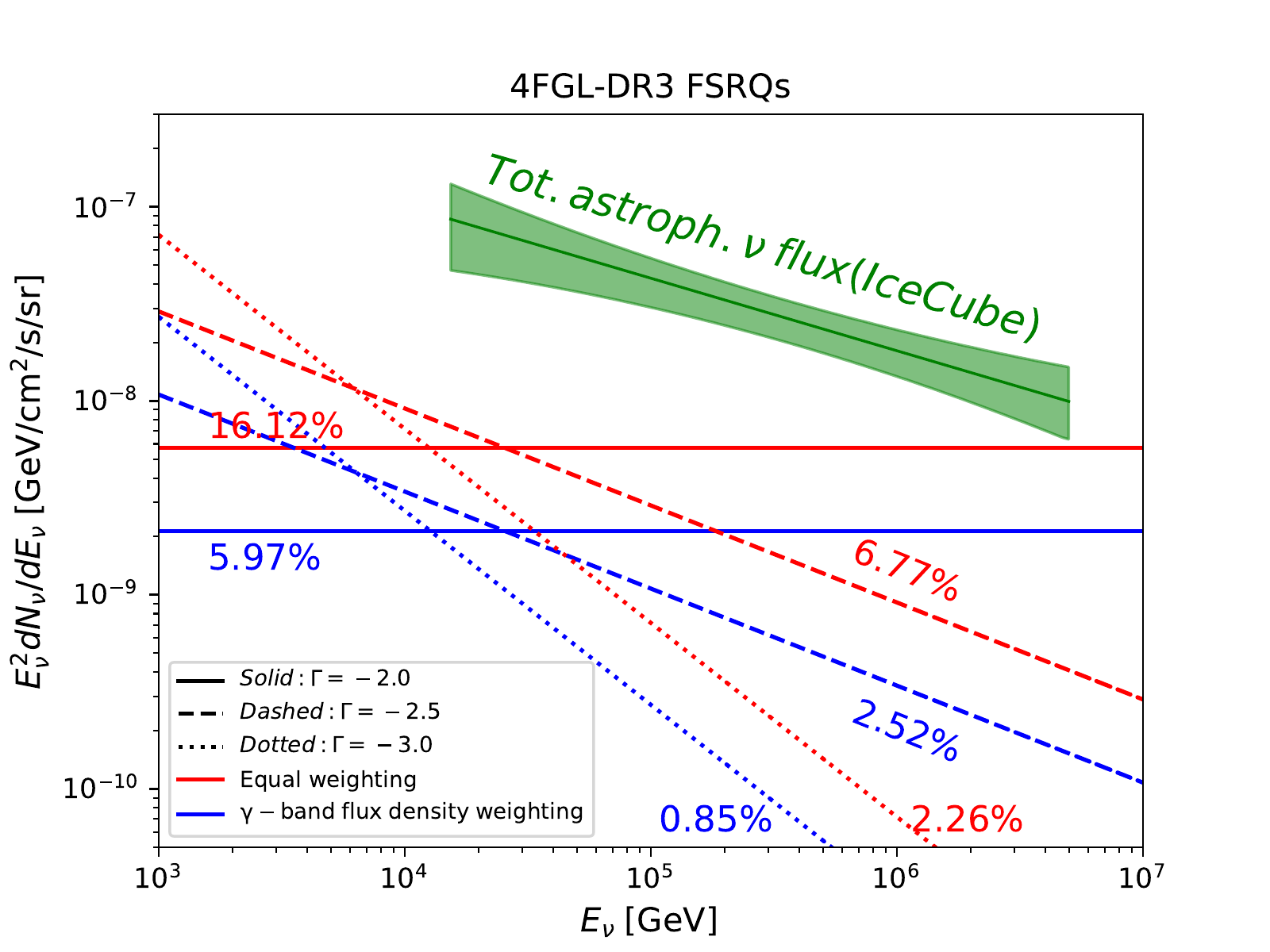}
\includegraphics[width=0.33\textwidth]{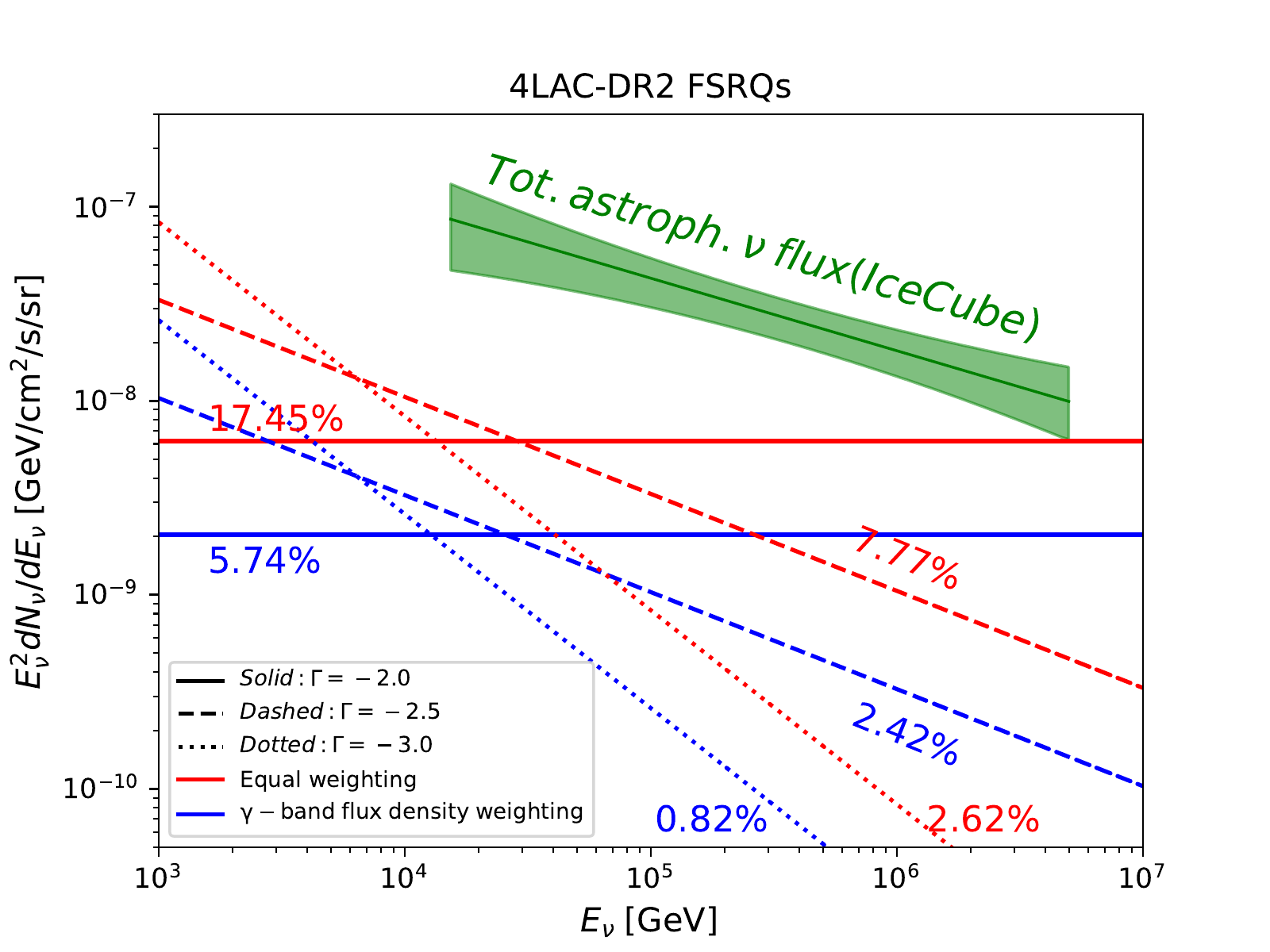}
\caption{The 95\% confidence level upper limits on the total neutrino flux from the sources in different samples. The constraints are compared to the all-sky diffuse neutrino flux measured by IceCube (green band). The number above each line displays the maximum fraction that the corresponding sample can contribute to the total diffuse neutrino flux. For each sample, we consider three different choices of the neutrino spectral index (solid, dashed and dotted lines) and two weighting schemes (red and blue lines).}
\end{figure*}

~

~

~

~

~

\newpage
~
\section{Sky map for the Fermi-LAT high-energy events}
\begin{figure*}[h]
\includegraphics[width=0.7\textwidth]{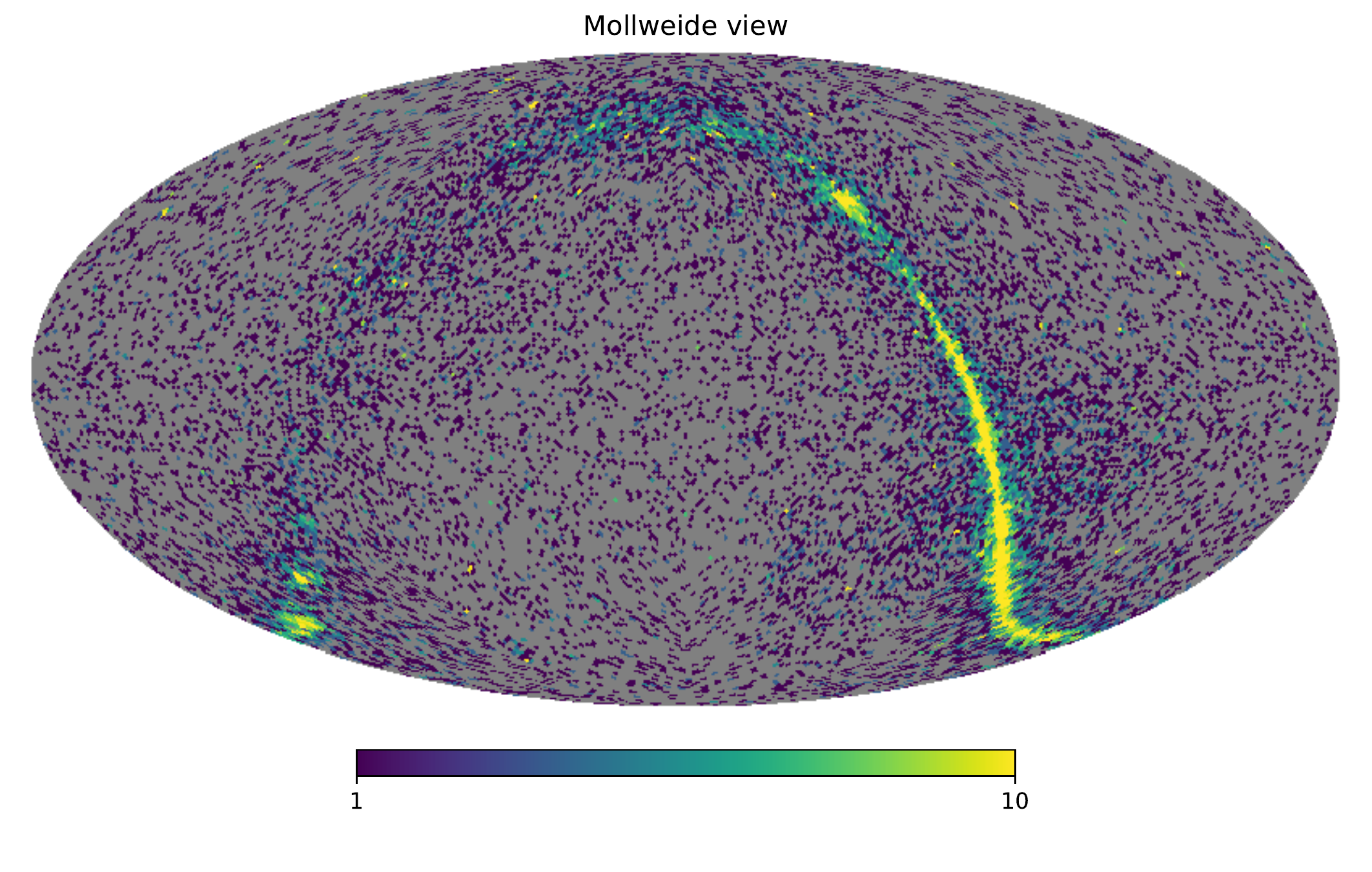}
\caption{Sky map for the Fermi-LAT high-energy ($>100\,{\rm GeV}$) events in equatorial coordinates. The Galactic plane of $|b|<10^\circ$ will be masked in our HEE analysis.}
\label{fig:heemap}
\end{figure*}
\end{document}